\documentclass[fleqn,usenatbib]{mnras}

\usepackage{newtxtext,newtxmath}

\usepackage[T1]{fontenc}

\DeclareRobustCommand{\VAN}[3]{#2}
\let\VANthebibliography\thebibliography
\def\thebibliography{\DeclareRobustCommand{\VAN}[3]{##3}\VANthebibliography}

\usepackage{graphicx}	
\usepackage{amsmath}	
\usepackage{soul}

\usepackage[dvipsnames]{xcolor}

\usepackage{mathtools}

\title[On dust evolution in binary discs -- theoretical and numerical modelling]{On dust evolution in planet-forming discs in binary systems. I -- Theoretical and numerical modelling: radial drift is faster in binary discs}

\author[F. Zagaria et al.]{
Francesco Zagaria$^{1,2,3,4}$\thanks{E-mail: fz258@cam.ac.uk},
Giovanni P. Rosotti$^{4,5}$,
Giuseppe Lodato$^{6}$
\\
$^{1}$Dipartimento di Fisica, Università degli Studi di Pavia, Via Agostino Bassi 6, I-27100 Pavia, Italy\\
$^{2}$Scuola Universitaria Superiore IUSS Pavia, Piazza della Vittoria 15, I-27100 Pavia, Italy\\
$^{3}$Institute of Astronomy, University of Cambridge, Madingley Road, Cambridge CB3 0HA, UK\\
$^{4}$Leiden Observatory, Leiden University, P.O.~Box 9513, NL-2300~RA Leiden, the Netherlands\\
$^{5}$School of Physics and Astronomy, University of Leicester, Leicester LE1 7RH, UK\\
$^{6}$Dipartimento di Fisica, Università degli Studi di Milano, Via Giovanni Celoria 16, I-20133 Milano, Italy
}

\date{Accepted XXX. Received YYY; in original form ZZZ}

\pubyear{2020}

\begin{document}
\label{firstpage}
\pagerange{\pageref{firstpage}--\pageref{lastpage}}
\maketitle

\begin{abstract}
    Many stars are in binaries or higher-order multiple stellar systems. Although in recent years a large number of binaries have been proven to host exoplanets, how planet formation proceeds in multiple stellar systems has not been studied much yet from the theoretical standpoint. In this paper we focus on the evolution of the dust grains in planet-forming discs in binaries. We take into account the dynamics of gas and dust in discs around each component of a binary system under the hypothesis that the evolution of the circumprimary and the circumsecondary discs is independent. It is known from previous studies that the secular evolution of the gas in binary discs is hastened due to the tidal interactions with their hosting stars. Here we prove that binarity affects dust dynamics too, possibly in a more dramatic way than the gas. In particular, the presence of a stellar companion significantly reduces the amount of solids retained in binary discs because of a faster, more efficient radial drift, ultimately shortening their lifetime. We prove that how rapidly discs disperse depends both on the binary separation, with discs in wider binaries living longer, and on the disc viscosity. Although the less-viscous discs lose high amounts of solids in the earliest stages of their evolution, they are dissipated slowly, while those with higher viscosities show an opposite behaviour. The faster radial migration of dust in binary discs has a striking impact on planet formation, which seems to be inhibited in this hostile environment, unless other disc substructures halt radial drift further in. We conclude that if planetesimal formation were viable in binary discs, this process would take place on very short time scales.
\end{abstract}

\begin{keywords}
    binaries: general -- circumstellar matter -- accretion, accretion discs -- protoplanetary discs -- planets and satellites: formation -- methods: numerical
\end{keywords}



\section{Introduction}
A significant fraction of stars are part of binary or higher-order multiple stellar systems (e.g., \citealt{Raghavan+10_2010ApJS..190....1R}). In addition, more than four thousand extra-solar planets around main-sequence stars have been discovered in the last twenty-five years (e.g., \citealt{Winn&Fabrycky15_2015ARA&A..53..409W}). 

Although traditionally the search for exoplanets has focused on single stars due to the technical difficulties of addressing multiple systems, in the last years more and more exoplanets have been detected in binaries (e.g., \citealt{Hatzes16_2016SSRv..205..267H,Martin18_2018haex.bookE.156M}). Some of those planets move around one of the two components of the system (e.g., \citealt{Hatzes+03_2003ApJ...599.1383H,Dumesque+12_2012Natur.491..207D}) and are known as S-type orbit planets. Others are circumbinary planets (e.g., \citealt{Doyle+11_2011Sci...333.1602D,Welsh+12_2012Natur.481..475W,Martin&Triaud14_2014A&A...570A..91M}) or planetary systems (as in the case of Kepler-47, e.g., \citealt{Orosz+19_2019AJ....157..174O}), and are known as P-type orbit ones. The number of exoplanet detections in S-type systems far outnumbers that of planets in P-type orbits: of the more than a hundred planets detected so far in binaries, only roughly a dozen are circumbinary (e.g., \citealt{Marzari&Thebault20_2019Galax...7...84M}). 

Such a growing census of exoplanets in multiple stellar systems was unexpected, given the number of works suggesting that planet formation around pre-main-sequence stars is inhibited by the presence of a stellar companion, 
both for S-type (e.g., \citealt{Thebault+06_2006Icar..183..193T,Thebault+08_2008MNRAS.388.1528T,Paardekooper+08_2008MNRAS.386..973P,Xie+10_2010ApJ...724.1153X,Fragner+11_2011A&A...528A..40F,Silsbee+15_2015ApJ...798...71S}) and P-type (e.g., \citealt{Marzari+13_2013A&A...553A..71M,Rafikov13_2013ApJ...774..144R,Lines+15_2015A&A...582A...5L,Lines+16_2016A&A...590A..62L}) orbit configurations. Nevertheless, such detections have shed a new interest on how planet formation proceeds around multiple stellar systems \citep{Marzari&Thebault20_2019Galax...7...84M}. 

There is general agreement that planet formation takes place in the proto-planetary discs, made of gas and dust, that orbit new-born stars. As proto-planetary discs are dynamical objects and contain the material reservoir out of which planets are assembled, studying their secular evolution is of paramount importance to gather new insights into how planets are formed. 

To this end, \citet{Jensen+94_1994ApJ...429L..29J,Jensen+96_1996ApJ...458..312J}, using the James Clerk Maxwell Telescope (JCMT), pioneered the observational studies of the (sub-)millimetre continuum dust emission in binary discs in Taurus, $\rho$~Ophiuchus and Upper~Scorpius. Since then, numerous surveys looking for proto-planetary discs in multiple stellar systems have been conducted, firstly with the Owens Valley Radio Observatory (OVRO) telescope and the Sub-Millimetre Array (SMA) and later with the Atacama Large Millimeter/sub-millimeter Array (ALMA) in the young ($\sim1-3\text{ Myr}$) Taurus \citep{Patience+08_2008ApJ...677..616P, Harris+12_2012ApJ...751..115H,Akeson&Jensen14_2014ApJ...784...62A,Akeson+19_2019ApJ...872..158A,Manara+19_2019A&A...628A..95M}, $\rho$ Ophiuchus \citep{Cox+17_2017ApJ...851...83C,Zurlo+20_2020MNRAS.496.5089Z} and Lupus \citep{Zurlo+20b_2020MNRAS.tmp.3477Z} star-forming regions as well as in the older ($\sim5-11\text{ Myr}$) Upper Scorpius OB-association \citep{Barenfeld+19_2019ApJ...878...45B}. In the youngest regions the observations proved that binary discs appear to be fainter and smaller in the (sub-)millimetre than their single-star analogues. Moreover, the disc continuum emission in each binary pair was shown to increase as a function of the system projected separation, matching single-star disc fluxes in wide binaries. Surprisingly, no similar trend was found in Upper Scorpius: in this region dust fluxes in binary and single-star discs are statistically indistinguishable. 

Several theoretical studies proved that binarity has a significant effect on planet-forming discs. In particular, due to tidal interactions between the disc and the companion star, binary discs are expected to be truncated at a fraction of the stellar separation. In some works disc truncation has been attributed to the resonant exchange of angular momentum at specific locations in the disc (e.g., \citealt{Goldreich&Tremaine79_1979ApJ...233..857G,Goldreich&Tremaine80_1980ApJ...241..425G,Artymowicz&Lubow94_1994ApJ...421..651A}), while in others it has been addressed to non-resonant mechanisms, such as perturbations in the gas density profile (e.g., \citealt{Papaloizou&Pringle77_1977MNRAS.181..441P}) or orbital crossing (e.g., \citealt{Paczynski77_1977ApJ...216..822P,Pichardo+05_2005MNRAS.359..521P}).

Tidal truncation has a profound impact on the secular evolution of planet-forming discs. This is because truncation effectively imposes a zero-flux outer boundary condition on the gas. Consequently, as the time scale of gas evolution is set by the viscous time scale at the disc outer edge \citep{Pringle81_1981ARA&A..19..137P}, it is expected that proto-planetary discs in binary systems evolve and disperse faster than single-star ones do. A number of observational studies aiming at constraining the fraction of disc-bearing stars in multiple stellar systems appear to confirm this trend both in the case of unresolved (e.g., \citealt{Cieza+09_2009ApJ...696L..84C,Kraus+12_2012ApJ...745...19K}) and resolved (e.g., \citealt{Daemgen+12_2012A&A...540A..46D,Daemgen+13_2013A&A...554A..43D}) discs.

Gas evolution in binaries in the presence of photo-evaporation has been investigated by \citet{Rosotti&Clarke18_2018MNRAS.473.5630R} and \citet{Alexander12_2012ApJ...757L..29A} in the case of circumstellar and circumbinary discs, respectively. They showed that significant differences emerge with respect to single-star discs due to the interplay between binary separation and photo-evaporation rate in determining proto-planetary disc lifetimes. In particular, \citet{Rosotti&Clarke18_2018MNRAS.473.5630R} suggested that in circumstellar discs in close binary systems the tidal truncation promotes outside-in instead of inside-out dispersal, as is in single-star discs, and hastens disc clearing. Moreover, they showed that mass accretion rates are expected to be higher in binary discs than in single-star ones.

Finally, the effects of binarity on the gas evolution and planet-formation processes were primarily studied using two- and three-dimensional simulations, such as in \citet{Muller&Kley12_2012A&A...539A..18M} and \citet{Picogna&Marzari13_2013A&A...556A.148P}. Those works showed that planet formation in close binaries is halted by the presence of a companion, e.g. due to spiral formation, higher disc temperatures, and generally a more hostile environment. Similar conclusions were obtained using one-dimensional steady-state models in \citet{Jang-Condell+08_2008ApJ...683L.191J} and \citet{Jang-Condell15_2015ApJ...799..147J}, considering the effects of solids parametrically (i.e., neglecting dust evolution).

Indeed, although most observational surveys have detected the dust continuum emission, no systematical study on the evolution of dust grains in discs in binary systems has been conducted to date. Notable exceptions are the work of \citet{Zsom+11_2011A&A...527A..10Z}, who took into account the growth and fragmentation of solids in the early stages of the evolution of a binary circumstellar disc, and some numerical studies focusing on circumbinary discs. Recent examples are \citet{Chachan+19_2019MNRAS.489.3896C} and \citet{Aly&Lodato20_2020MNRAS.492.3306A}, who examined dust trapping in pressure maxima and dust ring formation, respectively.

For single-star discs, it is known that dust dynamics is considerably different from that of the gas (e.g., \citealt{Testi+14_2014prpl.conf..339T}): dust grains are not subject to pressure, whose effect is responsible for the sub-keplerian motion of the gas. Friction due to this velocity difference makes the solids move radially inwards, in a process known as dust radial drift. From the theoretical point of view radial drift is expected to shape the disc structure with time, determining a sharp dust outer edge, as discussed in \citet{Birnstiel&Andrews+14_2014ApJ...780..153B}. Furthermore, because of radial drift, the extent of dust emission is expected to be more limited than the gaseous one, as shown in \citet{Rosotti+19a_2019MNRAS.486.4829R}. 

In the case of bright single-star discs in the young Lupus star-forming region, \citet{Ansdell+18_2018ApJ...859...21A} proved that the gas disc sizes exceed the dust ones by a factor of $\sim2$. Even more extreme ratios are possible, as in the peculiar case of CX Tau \citep{Facchini+19_2019A&A...626L...2F}, where the gas radius exceeds the dust one by a factor of $\sim4-5$. For binaries, we have explicit evidence of analogue results only in the case of the RW Aurigae system \citep{Rodriguez+18_2018ApJ...859..150R}. Finally, in \citet{Manara+19_2019A&A...628A..95M} it is shown that similar high ratios between gas and dust radii are needed to provide reasonable values of the eccentricities in the observed binaries.

In this paper we address the problem of dust secular evolution in binary systems, with particular interest in the role played by radial drift. As this last process is size-dependent, we employ the latest models of dust grain growth \citep{Birnstiel+12_2012A&A...539A.148B}. As for single-star discs, we follow the procedure of \citet{Rosotti+19b_2019MNRAS.486L..63R,Rosotti+19a_2019MNRAS.486.4829R}, who provided testable implications of their theoretical results, computing dust fluxes at ALMA wavelengths to be compared with observations. While in this paper we focus on the theoretical aspects, we refer the reader to a companion work for an exploration of the observational consequences of our models.

This paper is organised as follows. In Section~\ref{sec.2} we describe our model for the dynamics of proto-planetary discs in binary systems and introduce the code employed in our study. In Section~\ref{sec.3} we describe the theoretical predictions on gas and dust secular evolution in binary discs and test them in a representative case. Sections~\ref{sec.4} and \ref{sec.5} are devoted to the analysis of dust depletion in binaries as a function of the main parameters of the system. In Section~\ref{sec.6} we analyse the consequences of our evolution models on planet formation in binary systems and compare our results with those of similar works in the literature. Finally, in Section~\ref{sec.7} we draw our conclusions.

\section{Model description and numerical methods}\label{sec.2}
In this paper we follow the general approach of \citet{Rosotti&Clarke18_2018MNRAS.473.5630R} in modelling gas in proto-planetary discs in binary systems, without taking into account the effects of internal photo-evaporation. This choice is motivated in Section~\ref{sec.3}. Furthermore, as we are interested in the evolution of dust grains, we adapt the methods introduced in \citet{Rosotti+19b_2019MNRAS.486L..63R,Rosotti+19a_2019MNRAS.486.4829R} to the case of binaries.

We consider the evolution of the circumstellar discs in binaries, rather than the circumbinary discs. We assume such evolution to be independent of that of the companion disc(s). This approximation restricts the validity of our results to the phase of evolution of binary discs in which the re-supply of material from outside of their truncation radius can be neglected. In particular, this is the case of evolved Class II discs, which are our main modelling target.

The secular evolution of proto-planetary discs in binary systems is reproduced employing a one-dimensional finite-differences grid code. Gas dynamics is ruled by the viscous-diffusion equation, while for dust the code enforces the simplified treatment of grain growth described in \citet{Birnstiel+12_2012A&A...539A.148B}, coupled with the single-fluid advection equation introduced in \citet{Laibe&Price14_2014MNRAS.444.1940L} to model radial drift. In a companion paper we will estimate dust opacities at ALMA wavelengths in order to compute synthetic surface brightness profiles and dust fluxes to be compared with observations.

The architecture of the code we use is described in detail in \citet{Booth+17_2017MNRAS.469.3994B} and we refer the reader to this paper for any further explanations. As this code was written to study dusty single-star discs, some modifications are needed in order to correctly model the evolution of binary discs.

\paragraph*{Gas evolution} As in \citet{Rosotti&Clarke18_2018MNRAS.473.5630R}, instead of enforcing an explicit torque in the viscous-diffusion equation (e.g., \citealt{Lin&Papaloizou86_1986ApJ...309..846L}), we impose a zero-flux condition, solving the following closed-boundary problem:
\begin{equation}\label{eq.2.1}
    \begin{dcases}
        \dfrac{\partial\Sigma}{\partial t}=\dfrac{3}{R}\dfrac{\partial }{\partial R}\biggl[R^{1/2}\dfrac{\partial}{\partial R}\bigl(\nu\Sigma_\text{g}R^{1/2}\bigr)\biggr] & \text{if }R\leq R_\text{trunc},\\
        \dfrac{\partial}{\partial R}\bigl(\nu\Sigma_\text{g}R^{1/2}\bigr)=0 & \text{at }R=R_\text{trunc},
    \end{dcases}
\end{equation}
where $R$ is the cylindrical radius, $\Sigma$ and $\Sigma_\text{g}=\Sigma_\text{g}(R,t)$ are the total disc surface density and the gas surface density, respectively, $\nu$ is the disc viscosity and $R_\text{trunc}$ is the tidal truncation radius, which is taken as a free parameter. 

This choice is justified by the very steep radial dependence of the tidal torque, $\propto R^{-4}$ (e.g., \citealt{Lin&Papaloizou79_1979MNRAS.186..799L,Lin&Papaloizou86_1986ApJ...309..846L,Goldreich&Tremaine80_1980ApJ...241..425G}), which determines an abrupt fall of the gas surface density in a thin radial interval close to the truncation radius. Our approximation considers this interval to be infinitesimally small. In Appendix~\ref{app:1} we study the case of an explicit torque, discussing the difference from the current implementation and proving that the zero-flux assumption works properly.

As in several previous studies of evolved binary systems (e.g., \citealt{Bath&Pringle81_1981MNRAS.194..967B}), Eq.~\ref{eq.2.1} is solved on a grid equally spaced in $R^{1/2}$. In our case, the grid extends from $R_\text{in}=0.01\text{ au}$ to $R_\text{out}=R_\text{trunc}$ and is made up of 250 cells. In order to assess the effects of binarity on disc dynamics, we use different values of the tidal truncation radius, $R_\text{trunc}=25,\,50,\,100,\,150,\,250\text{ and }500\text{ au}$, corresponding to binary separations of $a=3\times R_\text{trunc}$ for equal mass stars, in circular orbits \citep{Papaloizou&Pringle77_1977MNRAS.181..441P}. For eccentric orbits or non-equal-mass binaries, the separation is larger for the same $R_\text{trunc}$ (e.g., \citealt{Artymowicz&Lubow94_1994ApJ...421..651A}).

\paragraph*{Dust evolution} The main quantity determining dust dynamics is the Stokes number, $\text{St}$ (e.g., \citealt{Whipple72_1972fpp..conf..211W,Weidenschilling77_1977MNRAS.180...57W}). In the Epstein regime, valid for small dust grains, the Stokes number is given by:
\begin{equation}\label{eq.2.2}
    \text{St}\sim\dfrac{\pi}{2}\dfrac{a}{\Sigma_\text{g}}\rho_\text{s},
\end{equation}
where $a$ is the dust grain size and $\rho_\text{s}$ is the bulk density of dust grains, which is set to be $\rho_\text{s}=1\,\text{g cm}^{-3}$. Dust particles with $\text{St}\ll$~1 are tightly coupled with gas, while those with $\text{St}\gg 1$ evolve independently of the gas and do not move radially. When $\text{St}\sim 1$ dust grains experience the fastest drift towards the central star.

The grain size is determined as in \citet{Birnstiel+12_2012A&A...539A.148B}. This prescription has the advantage of both reproducing the results of the full coagulation/fragmentation models (e.g., \citealt{Brauer+08_2008A&A...480..859B,Birnstiel+09_2009A&A...503L...5B,Birnstiel+10_2010A&A...513A..79B}) and being computationally less expensive. As described in \citet{Birnstiel+12_2012A&A...539A.148B}, at each disc radius two dust populations are evolved in time: a population of small grains, whose size is set to be the monomer grain size, $a_\text{min}=0.1\,\mu\text{m}$, and a population of large grains, whose size, $a_\text{max}$, is set by the combined effect of grain growth, fragmentation and radial drift. Two populations of solids with different sizes are enough to reproduce the results of models with a continuum of grain sizes, at a much smaller computational cost \citep{Birnstiel+12_2012A&A...539A.148B}. This is because most of the mass is in large grains, with a non-negligible fraction in very small grains in the fragmentation regime. Finally, it should be considered that the two-population model was benchmarked against single-star disc models and might not behave as well in the case of close binaries.

In those regions of the disc in which fragmentation is the main process limiting dust growth, the maximum grain size is set by:
\begin{equation}\label{eq.2.3}
    a_\text{frag} = f_\text{frag}\dfrac{2}{3\pi}\dfrac{\Sigma_\text{g}}{\rho_\text{s}\alpha}\dfrac{u_\text{f}^2}{c_\text{s}^2}.
\end{equation}
Here $f_\text{frag}=0.37$ is an order of unity factor set by the comparison with the full coagulation/fragmentation models \citep{Birnstiel+12_2012A&A...539A.148B}. $u_\text{f}$ is the dust fragmentation velocity, which is set to be $u_\text{f}=10\,\text{ m s}^{-1}$, valid for icy grains \citep{Gundlach&Blum15_2015ApJ...798...34G}. $\alpha$ is the \citet{Shakura&Sunyaev73_1973A&A....24..337S} prescription, which parametrically takes into account the effects of turbulent viscosity on the evolution of the gas. Finally, $c_\text{s}$ is the gas sound speed.

Conversely, in those regions of the disc in which radial drift is the main process limiting dust growth, the grain size is set by:
\begin{equation}\label{eq.2.4}
    a_\text{drift} = f_\text{drift}\dfrac{2}{\pi}\dfrac{\Sigma_\text{d}}{\rho_\text{s}}\dfrac{V_\text{K}^2}{c_\text{s}^2}\gamma^{-1},
\end{equation}
obtained equating dust drift and growth time scales (meaning that grains drift as fast as they grow). Here $f_\text{drift}=0.55$ is an order of unity factor set by the comparison with the full coagulation/fragmentation models \citep{Birnstiel+12_2012A&A...539A.148B}. $\Sigma_\text{d}=\Sigma_\text{d}(R,t)$ is the dust surface density, $V_\text{K}=\sqrt{GM_*/R}$ is the Keplerian velocity and $\gamma=\lvert d\log p/d\log R\lvert$ is the polytropic index, $p\propto\rho_\text{g}^\gamma$, where $p$ is the gas pressure and $\rho_\text{g}$ is the mid-plane gas surface density.

The maximum grain size is set as $a_\text{max}=\text{min}(a_\text{frag},a_\text{drift})$. In general, the inner region of the disc is fragmentation dominated, while the outer regions are in the drift-dominated regime. The main factor determining the relative importance of the two regimes is viscosity: at a given time, for larger values of $\alpha$, the fragmentation-dominated region expands to larger radii. The time-dependence in $a_\text{frag}$ and $a_\text{drift}$ is due to the (gas and dust) surface density profiles only. As $\Sigma_\text{d}<\Sigma_\text{g}$ and this inequality sharpens as time goes by, we expect that the drift-dominated region expands to smaller radii on secular time scales.

Finally, in the simplified treatment of dust introduced in \citet{Birnstiel+12_2012A&A...539A.148B}, each of the two populations has a fraction $f_\text{mass}$ of the total disc dust mass according to a factor set by the comparison with the full coagulation/fragmentation models \citep{Birnstiel+12_2012A&A...539A.148B}: $f_\text{mass}=0.97$, if $a_\text{max}=a_\text{drift}$, and $f_\text{mass}=0.75$, if $a_\text{max}=a_\text{frag}$. From now on, if not otherwise stated, we will refer to the grain size of the large population simply as the \textquotedblleft grain size\textquotedblright.

As described in \citet{Booth+17_2017MNRAS.469.3994B}, the code takes into account diffusion and radial drift of dust grains throughout the disc. The latter is achieved by computing the velocity of dust particles from their size and evolving the dust fraction according to the single-fluid model introduced in \citet{Laibe&Price14_2014MNRAS.444.1940L}. This approach has the advantage of taking into account both the effect of gas drag on dust particles and that of dusty grains on gas which can substantially affect disc evolution, e.g., in presence of local dust accumulations at pressure maxima (e.g., \citealt{Dipierro+18_2018MNRAS.479.4187D,Garate+20_2020A&A...635A.149G}). However, this is not expected to play a role in our case, in which, as we shall see, $\Sigma_\text{d}$ is always much smaller than $\Sigma_\text{g}$. 

As for gas, to take into account the effects of binarity also on dust evolution, a similar closed-outer-boundary condition has been enforced both in the diffusion and advection equations. As a matter of fact, this correction brings about negligible modifications to our final results. This happens because the outer regions of the disc retain only small dust grains, which are well-coupled with the gas. Those grains move along with the gas, which already takes into account the effect of binarity as prescribed in Eq.~\ref{eq.2.1}. Then, by construction, also the dust naturally does.

\paragraph*{Disc structure and initial conditions} As in \citet{Rosotti+19b_2019MNRAS.486L..63R,Rosotti+19a_2019MNRAS.486.4829R}, we set a locally isothermal disc with:
\begin{equation}\label{eq.2.5}
    T=88.23\,\biggl(\dfrac{R}{10\text{ au}}\biggr)^{-1/2}\text{ K}.
\end{equation}
This choice agrees with the  \citet{Chiang&Goldreich97_1997ApJ...490..368C} prescription for a solar mass star, $M_*=M_\odot$, and corresponds to a disc with aspect ratio of $H/R\sim1/30$ at $R=1\text{ au}$. Here $H$ is the disc scale-height, defined as $H=c_\text{s}/\Omega_\text{K}$, where $\Omega_\text{K}=\sqrt{GM_*/R^3}$ is the Keplerian angular velocity and $c_\text{s}=\sqrt{\mathcal{R}T/\mu}$, the locally isothermal gas sound speed, is defined in terms of the gas temperature profile, the gas mean molecular mass, $\mu=2.4$, and the gas constant, $\mathcal{R}$.

Some previous works that studied circumstellar disc evolution in binary systems suggested that higher mid-plane temperatures should be expected due to viscous heating, shock waves and mass transfer in their two- and three-dimensional simulations (e.g., \citealt{Nelson00_2000ApJ...537L..65N,Muller&Kley12_2012A&A...539A..18M,Picogna&Marzari13_2013A&A...556A.148P}). We will provide a detailed analysis of the consequences of this choice on our work later in this paper. As for now, we remark that those studies considered binary separations of $a\leq50\text{ au}$, by far smaller than ours. Furthermore, eq.~\ref{eq.2.5} and the temperature profiles in \citet{Nelson00_2000ApJ...537L..65N} and \citet{Picogna&Marzari13_2013A&A...556A.148P} are substantially different only in the innermost disc region, with the reference temperature at 10~au being less than a factor of 2 higher in those studies and decreasing more steeply in the outer disc than ours as the stellar irradiation is not taken into account. As we are interested in the modelling the outer disc regions eq.~\ref{eq.2.5} seems a valid assumption.

Once the temperature profile has been set, it is possible to compute the disc viscosity at each radius through the \citet{Shakura&Sunyaev73_1973A&A....24..337S} $\alpha$ prescription, under the hypothesis that it only affects gas dynamics, as $\nu=\alpha c_\text{s}H$. For our chosen temperature profile, viscosity scales linearly with the disc radius: $\nu\propto R$. We have explored a range of viscosities encompassing the typical values expected in proto-stellar discs: $\alpha=10^{-4},\,10^{-3}$ and $10^{-2}$. From the observational point of view, they are motivated by several works (e.g., \citealt{Trapman+20_2020arXiv200511330T,Andrews20_2020arXiv200105007A}). We note that there is recent mounting evidence that the value $\alpha=10^{-2}$ might be too large to be compatible with observations (but see also \citealt{Lodato+17_2017MNRAS.472.4700L} and the recent work of \citealt{Flaherty+20_2020ApJ...895..109F} for the detection of turbulent motions in DM Tau corresponding to $\alpha\sim0.08$). Nevertheless, it is instructive to consider it because in this case fragmentation limits the grain size. As in \citet{Rosotti+19a_2019MNRAS.486.4829R}, we also consider the case $\alpha=0.025$, which, despite being only slightly higher than $\alpha=10^{-2}$, displays a significantly different behaviour.

Our initial condition for the disc surface density, $\Sigma=\Sigma_\text{g}+\Sigma_\text{d}$, is a self-similar profile (e.g., \citealt{Lynden-Bell&Pringle74_1974MNRAS.168..603L}):
\begin{equation}\label{eq.2.6}
    \Sigma(R,t=0) = \dfrac{M_0}{2\pi R_0R}\exp\left(-\dfrac{R}{R_0}\right).
\end{equation}
Here $M_0$ is the initial disc mass, which is set to be $M_0=0.1M_\odot$ and $R_0$ is a characteristic scale radius, defined as the radius enclosing 63 per cent of the initial disc mass. Also in the case of the scale radius we employ different initial parameters in our simulations, corresponding to $R_0=10,\,30,\text{ and }80\text{ au}$. The initial dust mass and surface density are determined by the choice of a uniform initial dust fraction: $\epsilon=0.01$.

\section{Binarity affects gas and dust evolution}\label{sec.3}

\subsection{Theoretical expectations}

\paragraph*{Gas evolution} As outlined in \citet{Rosotti&Clarke18_2018MNRAS.473.5630R}, the secular evolution of gas in binary discs can be divided into three consecutive stages:

\begin{itemize}
    \item Initially the gas viscously expands outwards following the typical similarity solution which applies to single-star discs (e.g., \citealt{Lynden-Bell&Pringle74_1974MNRAS.168..603L}):
    \begin{equation}\label{eq.2.8}
        \Sigma_\text{g}(R,t)=\dfrac{M_0}{2\pi R_0R}T^{-3/2}\exp\left(-\dfrac{R}{R_0T}\right),
    \end{equation}
    where $T=1+t/t_\nu$ is a dimensionless time variable in which $t_\nu=$~$t_\nu(R=R_0)$ is the disc viscous time scale computed at the initial scale radius. Here $t_\nu$ is defined as (e.g., \citealt{Lynden-Bell&Pringle74_1974MNRAS.168..603L}):
    \begin{equation}\label{eq.2.7}
        t_\nu=\dfrac{R^2}{3\nu}=\dfrac{R^2}{3\alpha c_\text{s}H}, 
    \end{equation} 
    which is a decreasing function of $\alpha$.
    
    The gas becomes sensitive to the torque of the companion as soon as a relevant fraction of its mass reaches the outer boundary, $R_\text{out}=R_\text{trunc}$. The time-span, $t_\text{boundary}$, of this first stage is set by both the initial scale radius, which determines the fraction of the disc mass initially in the outer regions of the disc, and its viscous time scale at the truncation radius, $t_\nu(R=R_\text{trunc})$. As a consequence, we expect initially more compact (smaller $R_0$), less viscous (smaller $\alpha$) discs with a larger truncation radius to behave essentially as single-star discs, since $t_\text{boundary}$ would be longer.
    
    \item Once a relevant fraction of the gas has reached the disc outer radius, the gas surface density goes through a phase of re-adjustment, departing from the similarity solution in Eq.~\ref{eq.2.8}. At the end of this process, the gas surface density is described by the similarity solution that solves Eq.~\ref{eq.2.1} \citep{Rosotti&Clarke18_2018MNRAS.473.5630R}:
    \begin{equation}\label{eq.2.9}
        \Sigma_\text{g}(R,t)=\dfrac{M_0}{8R_\text{trunc}^{1/2}R^{3/2}}\sin\left(\dfrac{\pi R^{1/2}}{2R_\text{trunc}^{1/2}}\right)\exp\left(-\dfrac{t}{t_\text{trunc}}\right),
    \end{equation}
    where $t_\text{trunc}=16R_\text{trunc}^2/3\pi^2\nu_\text{trunc}$, in which $\nu_\text{trunc}=\nu(R=R_\text{trunc})$, and is valid only for $R\leq R_\text{trunc}$. We note that Eq.~\ref{eq.2.8} and Eq.~\ref{eq.2.9} share the same radial dependence at small radii, $\Sigma_\text{g}\propto R^{-1}$, while at larger radii $\Sigma_\text{g}\propto R^{-3/2}$ in binary discs. 

    \item Finally, the secular evolution proceeds on a time scale set by $t_\nu(R=R_\text{trunc})$. This time scale is much shorter than that of a single-star disc, which, instead, is set by $t_\nu(R=R_\text{out})$. Here $R_\text{out}$ is the disc outer radius, which viscously expands with time. For this reason, we expect the gas to evolve faster in the case of binaries. 
\end{itemize}

This last result is confirmed by looking at the gas mass accretion rate onto the central star. In the single-star case (e.g., \citealt{Lynden-Bell&Pringle74_1974MNRAS.168..603L}):
\begin{equation}\label{eq.2.10}
    \dot{M}=\dfrac{1}{2}\dfrac{M_0}{t_\nu}T^{-3/2},
\end{equation}
where $t_\nu=t_\nu(R=R_0)$ is the viscous time scale computed at the initial scale radius. On the other hand, once self-adjustment occurred, in binary discs the mass accretion rate can be computed as \citep{Rosotti&Clarke18_2018MNRAS.473.5630R}:
\begin{equation}\label{eq.2.11}
    \dot{M}=\dfrac{M_0}{t_\text{trunc}}\exp\left(-\dfrac{t}{t_\text{trunc}}\right).
\end{equation}
Clearly, while in Eq.~\ref{eq.2.10} the mass accretion rate declines as a power-law with time, in Eq.~\ref{eq.2.11} it has an exponential decrease, which sets the total lifetime of a binary disc to be $\sim 2-3$ times $t_\text{trunc}$.

In \citet{Rosotti&Clarke18_2018MNRAS.473.5630R}  internal X-ray photo-evaporation is also taken into account, showing that it is relevant on gas evolution only if binary discs are large enough for gap opening to take place. This is the case of almost all our models, with the possible exception of the $R_\text{trunc}=25\text{ au}$ case, where clearing would take place outside-in instead of inside-out (see Fig.~3 in \citealt{Rosotti&Clarke18_2018MNRAS.473.5630R}). Photo-evaporation would hasten disc dispersal with respect to the pure viscous scenario. Indeed, the time scale of gas depletion within the wind-opened gap would be set by the viscous time scale at the gap edge, which is smaller than $t_\nu(R=R_\text{trunc})$. Moreover, because of the steep decrease of the gas surface density in the outer disc, $\Sigma_\text{g}\propto R^{-3/2}$, disc clearing due to photo-evaporation would be faster than the pure viscous draining. 

It is known that a two-time-scale evolution framework is necessary to explain disc dispersal on secular time scales (e.g. \citealt{Haisch+01_2001ApJ...553L.153H,Ribas+14_2014A&A...561A..54R,Ercolano&Pascucci17_2017RSOS....470114E}). Nevertheless we decided not to consider photo-evaporation as it would make even less longer-lived our already short-lifetime models (see e.g., Section~\ref{sec.4}).

\paragraph*{Dust evolution} Also for the dust we expect a similarly faster evolution in binaries as in freely expanding discs. To see this, consider the extreme case in which dust is perfectly coupled with the gas. In this case, it will naturally experience the same faster evolution of gas molecules. 

In addition, in the case of binaries we expect radial drift to be a more efficient mechanism to remove dust grains. In single-star discs, radial drift acts removing large and fast-drifting grains, yet it leaves behind smaller grains well-coupled with the gas. These grains are entrained in viscous expansion and move outwards \citep{Rosotti+19a_2019MNRAS.486.4829R}. Because of this process, a reservoir of small, slowly growing grains is created in the outer regions of the disc, preventing an abrupt fall of the dust fraction with time. In the case of binaries this process cannot take place due to the presence of the closed outer boundary. As a consequence, no reservoir of solids acts to prevent a substantial dust depletion. 

Finally, even if the dust fraction were the same in binary and single-star discs, the time scale of dust growth would be different. The time scale of dust growth is defined as \citep{Birnstiel+12_2012A&A...539A.148B}:
\begin{equation}\label{eq.2.12}
    t_\text{growth}=\dfrac{1}{\Omega_\text{K}\epsilon}\propto\dfrac{R^{3/2}}{\epsilon},
\end{equation}
where $\epsilon$ is the dust-to-gas ratio. As for the viscous time scale, $t_\text{growth}(R=R_\text{out})$ is shorter in the case of binaries, where $R_\text{out}=R_\text{trunc}$, than in single-star discs, where $R_\text{out}$ increases with time. Because of this effect, dust grains in binaries are expected to reach the critical size at which they are removed by radial drift sooner than in the single-star disc case. 

\citet{Rosotti&Clarke18_2018MNRAS.473.5630R} considered the effects of internal photo-evaporation focusing on mass accretion rates and near infrared emission, none of which are influenced by the dust. Moreover, even though photo-evaporation is expected to hasten disc dispersal with respect to the pure viscous scenario, it would eventually take place only at late times, when the bulk of the dust has been already removed from the disc. In this case only small grains would be retained, which can be entertained in the wind (e.g., \citealt{Booth&Clarke21_2021MNRAS.502.1569B}).  Those reasons suggests to consider the effects of disc winds on dust in binary evolution in a future work. \\*

To sum up, we expect that the tidal effects due to the presence of a stellar companion will dramatically shorten the lifetime of both gas and dust in binary discs.

\subsection{A model case for gas and dust evolution in binaries}
It is time to test our theoretical expectations with numerical simulations. We compare the time-evolution of gas and dust in a freely expanding disc, with $R_0=10\text{ au}$ and $\alpha=10^{-3}$, and a binary disc, with $R_0=R_\text{trunc}=10\text{ au}$ and the same viscosity. This choice is motivated by the ease in distinguishing the single- and binary-disc behaviour as explained in the following paragraphs. In the next Sections we will examine the dependence of our results on the initial conditions.

\begin{figure}
    \centering
	\includegraphics[width=\columnwidth]{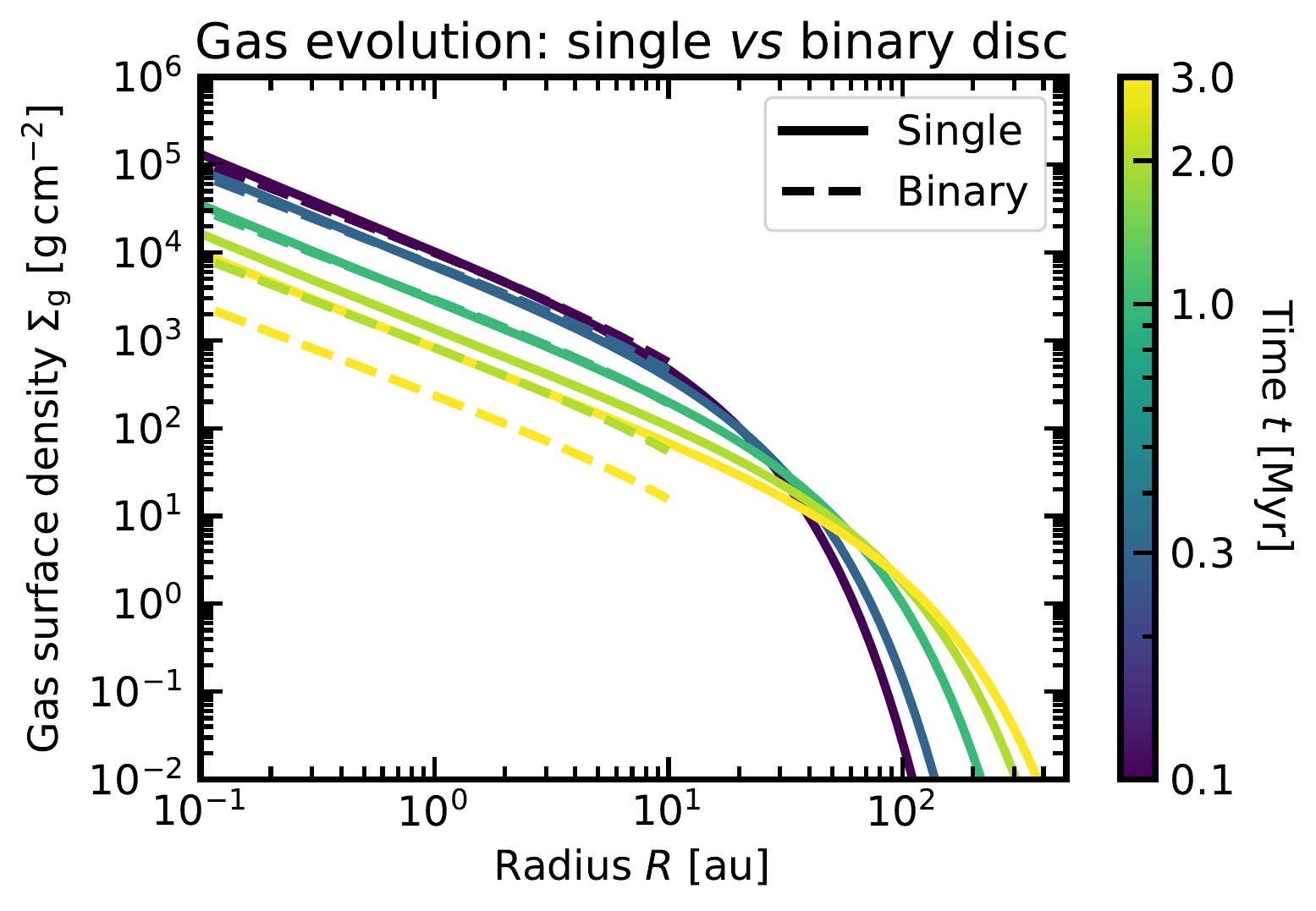}
    \caption{Gas surface density evolution for the freely expanding disc - solid lines - and binary disc - dashed lines - model reference test run with $\alpha=10^{-3}$ and $R_0=R_\text{trunc}=10\text{ au}$. The profiles are evaluated at $t=0.1,\,0.3,\,1,\,2\text{ and }3\text{ Myr}$.}
    \label{fig.3.1}
\end{figure}

\paragraph*{Gas evolution} First of all, we assess the effect of binarity on gas evolution. Fig.~\ref{fig.3.1} shows the radial dependence of the gas surface density in our model discs after $t=0.1,\,0.3,\,1,\,2\text{ and }3\text{ Myr}$. Despite having the same initial condition, as time goes by the gas surface density in the binary model falls below its single-star disc analogue. This can be seen in Fig.~\ref{fig.3.1} after $t\sim1\text{ Myr}$. Indeed, because of our choice of the model parameters, both discs initially have the same viscous time scale, $t_\nu(R=R_0)$. As time goes on the binary disc still evolves on this same time scale. On the contrary, due to the disc viscous expansion, such a time scale increases in the single-star disc case, determining a slower evolution than in the binary disc one.

\begin{figure*}
    \centering
	\includegraphics[width=\textwidth]{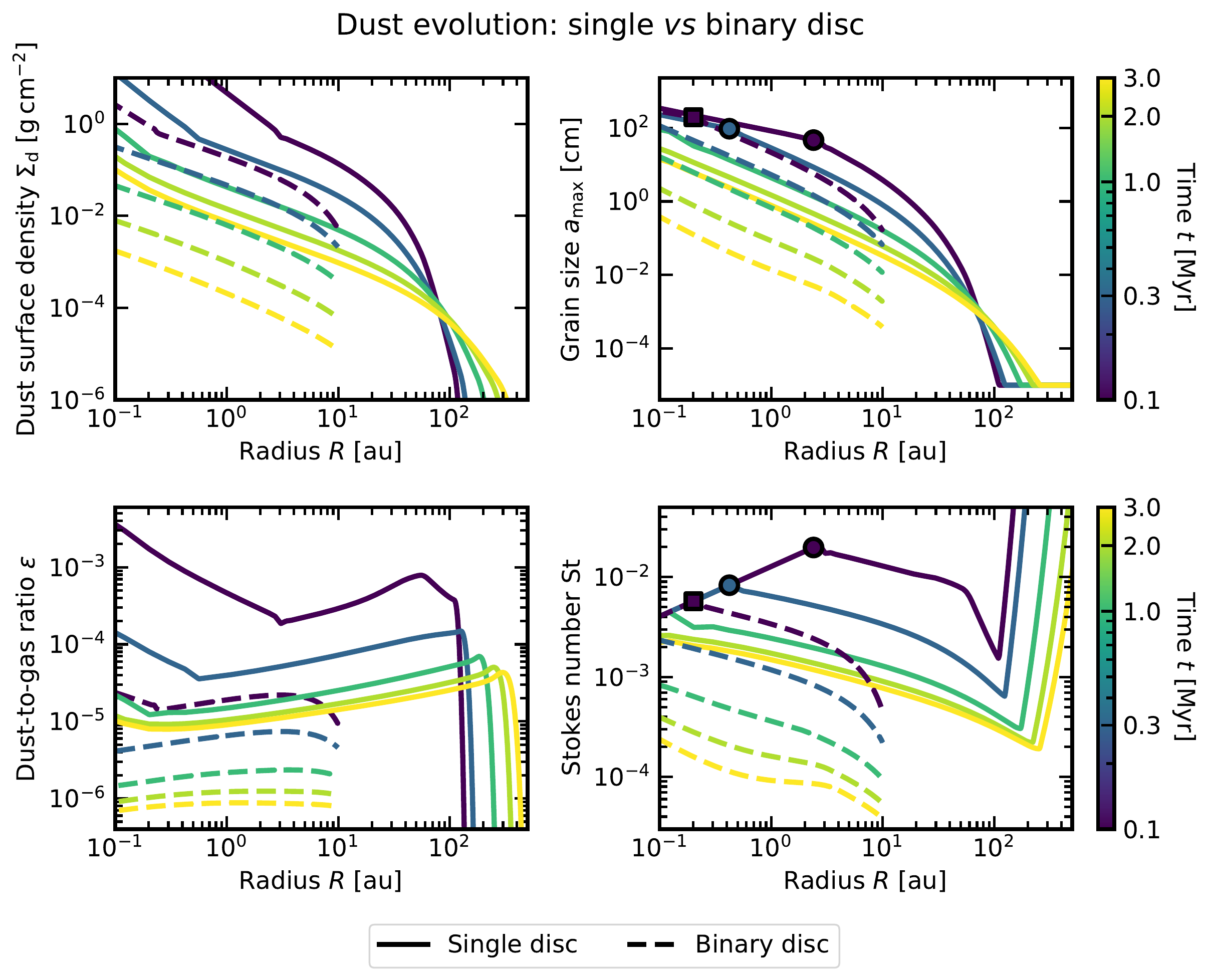}
    \caption{\textbf{Moving clockwise from the upper-left panel:} Evolution of the dust surface density, $\Sigma_\text{d}$, the maximum dust grain size, $a_\text{max}$, the Stokes number, $\text{St}$, and the dust-to-gas ratio, $\epsilon$, for our freely expanding disc - solid line - and binary disc - dashed line - model reference test run with $\alpha=10^{-3}$ and $R_0=R_\text{trunc}=10\text{ au}$. The profiles are evaluated at $t=0.1,\,0.3,\,1,\,2\text{ and }3\text{ Myr}$.}
    \label{fig.3.2}
\end{figure*}

To sum up, our theoretical expectations in the case of gaseous discs in binary systems are supported by the simulations: the gas depletes on a faster time scale than around single-stars.

\paragraph*{Dust evolution} Let us now focus on Fig.~\ref{fig.3.2}. Starting from the upper-left panel and moving clockwise, it displays the dust surface density, $\Sigma_\text{d}$, the maximum grain size, $a_\text{max}$, the Stokes number, $\text{St}$, and the dust-to-gas ratio, $\epsilon$, as a function of the disc radius for our single - solid lines - and binary - dashed lines - model discs, after $t=0.1,\,0.3,\,1,\,2\text{ and }3\text{ Myr}$. 

At first glance, the evolution of the dust in our binary model closely follows the main features of the single-star disc case already identified in \citet{Rosotti+19a_2019MNRAS.486.4829R}. However, some crucial differences need to be outlined. As can be seen from the time decay of the dust-to-gas ratio in the lower-left panel of Fig.~\ref{fig.3.2}, in both models the dust is depleted preferentially with respect to the gas. However, in the binary disc dust depletion is more dramatic than in the single-star model, with $\epsilon$ dropping below $10^{-6}$ within a few Myr. 

As outlined in \citet{Rosotti+19a_2019MNRAS.486.4829R}, in the single-star model we expect the dust grain size to attain two different regimes. In the inner regions of the disc the maximum grain size is limited by fragmentation, while in the outer regions it is set by radial drift. The transition between these two behaviours can be recognised as a knee in the radial profile of the grain size or of the Stokes number, as the upper-right and lower-right panels of Fig.~\ref{fig.3.2} show with the dots and the squares in the case of the single and binary disc model, respectively. As time goes on, as a consequence of dust and gas depletion, the maximum grain size decreases. Moreover, being dust preferentially depleted with respect to the gas, $a_\text{drift}$ decreases faster than $a_\text{frag}$ and the drift-dominated regime encompasses larger and larger regions of the disc.

Remarkably, in our binary model the dust surface density decreases so fast that the disc is fragmentation dominated only in the innermost region, in the earliest stage of its evolution. Apart from this case, all particles at any radius are large enough to experience radial drift and no reservoir of $\mu\text{m}$-size grains prevents the fast dispersal of dusty binary discs. This explains why the dust-to-gas ratio in the binary disc falls faster than in the single-star model as can be seen in the bottom-left panel of Fig.~\ref{fig.3.2}. As in the single-star case, as time goes by and the dust surface density decreases, radial drift becomes more and more selective, depleting the disc from smaller and smaller particles, ultimately reducing the size of the remaining dust grains. 

Even though not encompassing secular disc evolutionary time scales, similar results were also obtained by \citet{Zsom+11_2011A&A...527A..10Z}, who showed that both the grain size and the stopping time are reduced in binary discs by factors comparable with those in the earliest stage of our models. In qualitative agreement with the faster dust evolution in binaries is also the work of \citet{Panic+20_2020arXiv201207901P}, who show that the smallest circumprimary disc in their sample has the lowest mm flux, despite being youngest.

Finally, while in the single-star model the dust surface density shows a sharp outer edge at any time, no similar feature is present in the binary model because of the closed-outer-boundary condition imposed on both gas and dust evolution. In Appendix \ref{app:1} we will discuss how this feature changes if an explicit torque rather than the closed-boundary condition is imposed on gas evolution.

To sum up, as far as dust evolution is concerned, both single-star and binary discs share the same general features. However, in the binary case, radial drift reduces the dust-to-gas ratio faster than in the freely expanding model.

\section{Dependence of dust depletion on $\alpha$, $R_0$ and the truncation radius}\label{sec.4}
In this Section we discuss the dependence of our results on the disc parameters and we quantify how much dust the disc retains. In particular, we examine the variation of the integrated dust-to-gas ratio, $M_\text{dust}/M_\text{gas}$, with the viscous parameter, $\alpha$, the initial disc scale radius, $R_0$, and the tidal truncation radius, $R_\text{trunc}$.

\begin{figure*}
    \centering
    \includegraphics[width=0.99675\textwidth]{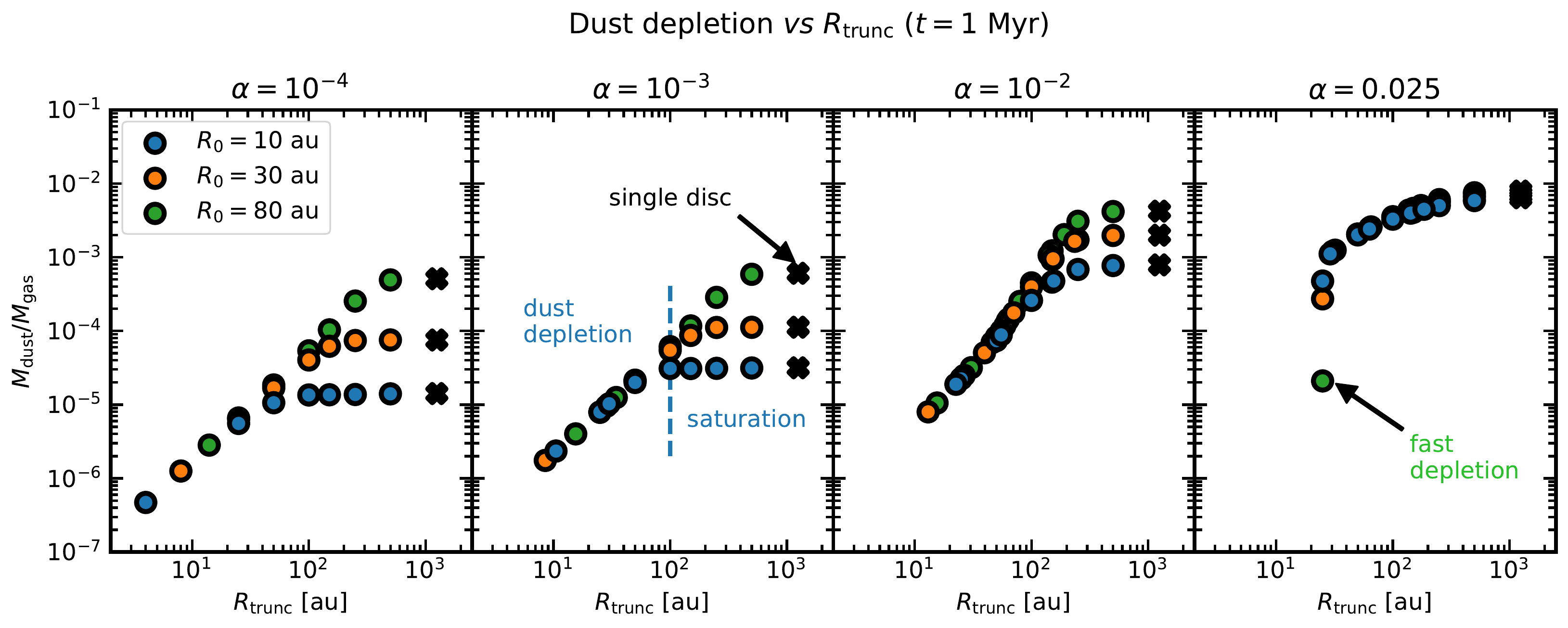} 
    \includegraphics[width=0.99675\textwidth]{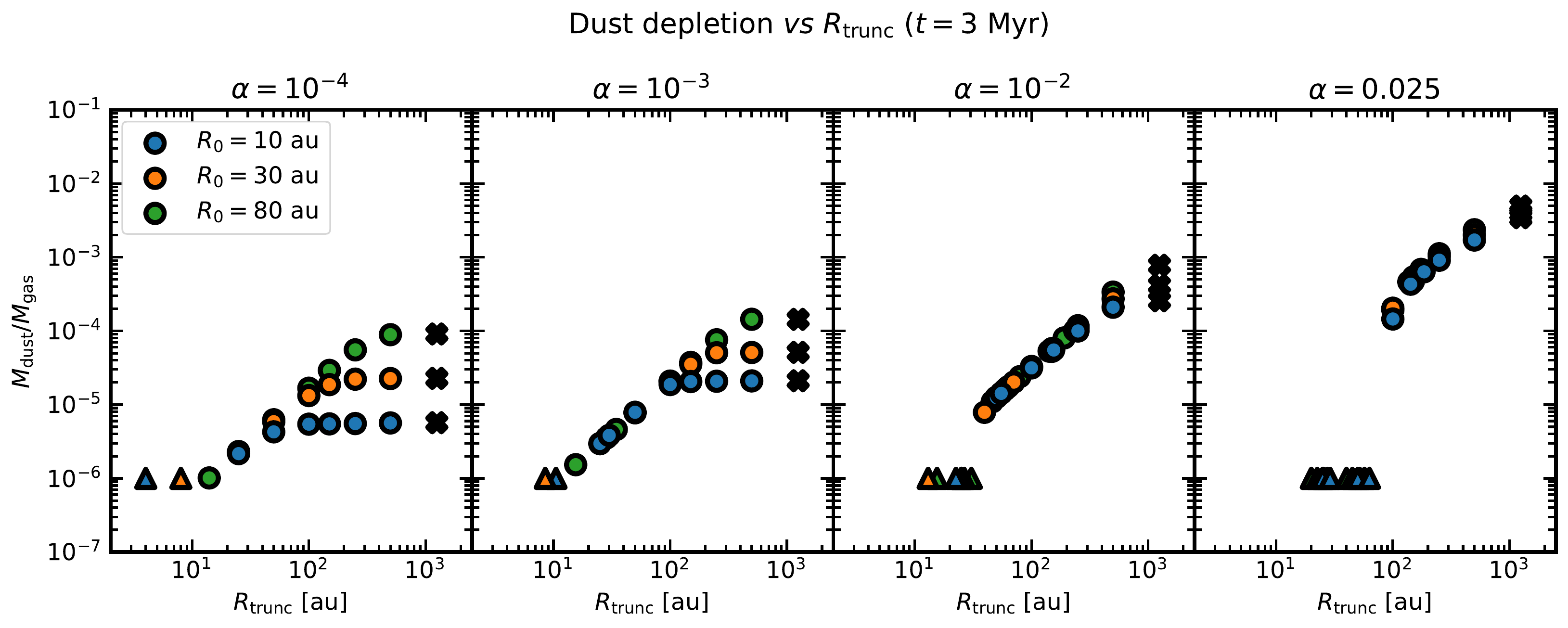}
    \caption{\textbf{Upper panel:} The integrated dust-to-gas ratio, $M_\text{dust}/M_\text{gas}$, is plotted as a function of the tidal truncation radius, $R_\text{trunc}$, for different values of $\alpha$, and the initial disc scale radius, $R_0$, after $t=1\text{ Myr}$. The black crosses indicate the value of $M_\text{dust}/M_\text{gas}$ corresponding to single-star discs with the same $\alpha$ and $R_0$. \textbf{Lower panel}: Same as in the upper panel but after $t=3\text{ Myr}$. The triangles identify upper limits for the integrated dust-to-gas ratio, once a sensitivity threshold has been set to $M_\text{dust}/M_\text{gas}=10^{-6}$.}
    \label{fig.4.1}
\end{figure*}

The upper panel in Fig.~\ref{fig.4.1} shows the dependence of $M_\text{dust}/M_\text{gas}$ on the disc tidal truncation radius, for different values of the viscosity and the initial disc scale radius, after $t=1\text{ Myr}$. The black crosses indicate the value of $M_\text{dust}/M_\text{gas}$ corresponding to single-star discs with the same parameters.

As a natural extension of the worked example in Section~\ref{sec.3}, let us focus on the $\alpha=10^{-3}$ and $R_0=10\text{ au}$ case and analyse the dependence of the integrated dust-to-gas ratio on $R_\text{trunc}$. As long as $R_\text{trunc}\geq100\text{ au}$, the binary disc is indistinguishable from a single-star disc, while dust gets progressively more depleted when $R_\text{trunc}$ approaches $R_0$. We can hypothesise that the truncation radius at which saturation occurs is such that the viscous time scale at that radius, $t_\nu(R_\text{trunc})$, equals the age of the disc. However, a direct computation shows that $t_\nu(R=100\text{ au})\sim5\text{ Myr}\gg1\text{ Myr}$. Another possibility to explain the dependence of the integrated dust-to-gas ratio with $R_\text{trunc}$ is that saturation occurs when the dust growth time scale at the truncation radius, $t_\text{growth}(R_\text{trunc})$, equals the age of the disc. However, as in the previous case, $t_\text{growth}(R=100\text{ au})\sim3\text{ Myr}\gg1\text{ Myr}$.

In Section~\ref{sec.3} we pointed out that the closed-outer-boundary condition starts to play a role in the evolution of binary discs only after the time $t_\text{boundary}$, in which a relevant fraction of the disc mass reaches the outer regions of the disc. We confirm that after $t=1\text{ Myr}$ only a tiny fraction of the disc mass went beyond $R=100\text{ au}$. For this reason, binary discs with $R_\text{trunc}>100\text{ au}$ essentially behave as single-star discs \textit{so far}. To put it in other words, $t_\text{boundary}\leq 1\text{ Myr}$ only for those binary discs with $R_\text{trunc}\leq 100\text{ au}$, showing that $t_\text{boundary}$ increases with $R_\text{trunc}$ as expected from our theoretical considerations in Section~\ref{sec.3}.

\paragraph*{Dependence on the initial scale radius} Let us now study how the initial disc scale radius $R_0$ affects the dependence of $M_\text{dust}/M_\text{gas}$ on the tidal truncation radius. From the upper panel in Fig.~\ref{fig.4.1}, a similar behaviour as in the case $R_0=10\text{ au}$ can be seen for $R_0=30\text{ au}$, with $M_\text{dust}/M_\text{gas}$ increasing with $R_\text{trunc}$ and then levelling out at $R_\text{trunc}=$~$250\text{ au}$, and for $R_0=80\text{ au}$, where saturation of the integrated dust-to-gas ratio occurs at $R_\text{trunc}=500\text{ au}$. The fact that the saturation radius moves to larger values as the initial disc scale radius increases can be explained using the definition of $R_0$ as the radius containing 63 per cent of the initial disc mass. Discs with a larger initial scale radius have more massive outer regions than discs with a smaller one, suggesting that, at a given stage of their evolution, a larger fraction of the disc mass can interact with the outer boundary. This is equivalent to say that $t_\text{boundary}$ is a decreasing function of $R_0$, as it was expected from our theoretical considerations in Section~\ref{sec.3}. 

The last feature to be clarified is why discs with larger $R_0$ retain more dust than the homologous with a smaller initial disc scale radius. As can be seen from Fig.~\ref{fig.4.1}, this is also found in single-star discs. In this case, initially more extended discs, being more massive in the outer regions, are also richer in dust at large radii than initially smaller ones. In the outer disc, radial drift and grain growth have longer time scales, implying that less dust is lost onto the star. Additionally, part of the dust is transported outwards by the gas viscous expansion, further slowing the evolutionary time scale. As a consequence of both these effects, if $R_0$ is bigger, more dust is stored in the outer disc and shielded against radial drift. 

Due to the presence of a closed outer boundary at $R=R_\text{trunc}$, no outward transport of dust is possible in binary discs. However, the same considerations on the time scale of dust growth apply, explaining the increase of $M_\text{dust}/M_\text{gas}$ with $R_0$ in binaries, too. Additionally, it can be seen from the upper panel in Fig.~\ref{fig.4.1} that, for small values of $R_\text{trunc}$, the initial disc scale radius does not affect the integrated dust-to-gas ratio. This happens as the exponential tail in the initial condition (see Eq.~\ref{eq.2.6}) is missing. This is consistent with the fact that when binarity is important the secular evolution of the gas is described by Eq.~\ref{eq.2.9} where $R_0$ does not play a role.

On a related note, we remark that considering models with $R_\text{trunc}\leq R_0$ is equivalent to assuming an initial power-law surface density (without exponential drop). We took into account such a scenario so as to explore an alternative initial condition.

\paragraph*{Dependence on the disc viscosity} Let us now explore the effects of changing the value of the disc viscosity, $\alpha$. From the upper panel in Fig.~\ref{fig.4.1} it is clear that for higher values of $\alpha$ more dust is retained both in binary discs and in single-star ones: the integrated dust-to-gas ratio increases. 

As shown in the upper panel of Fig.~\ref{fig.4.1}, for $\alpha=10^{-4}$ the integrated dust-to-gas ratio undergoes a similar behaviour with $R_\text{trunc}$ as for $\alpha=10^{-3}$, for each initial disc scale radius $R_0$. Saturation of $M_\text{dust}/M_\text{gas}$ also occurs at similar values of $R_\text{trunc}$, but the integrated dust-to-gas ratio at the plateau is slightly smaller than in the $\alpha=10^{-3}$ case. Remarkably, the increased depletion affects especially the case of $R_0=10\text{ au}$. We will provide an explanation of this feature in Section~\ref{sec.5}.

Although the general picture is the same also for $\alpha=10^{-2}$, some important differences occur. The integrated dust-to-gas ratio has a steeper dependence on $R_\text{trunc}$ than for $\alpha=10^{-3}$ and no saturation arises until $R_\text{trunc}=500\text{ au}$, regardless of $R_0$. The fact that the truncation radius at which $M_\text{dust}/M_\text{gas}$ levels out increases can be accounted for noticing that the viscous time scale is a factor of 10 lower than in the $\alpha=10^{-3}$ case. Indeed, after $t=1\text{ Myr}$, more viscous discs are comparably more evolved and a larger fraction of their mass has reached the outer radius. In other words, $t_\text{boundary}$ decreases with $\alpha$ as expected from our theoretical considerations in Section~\ref{sec.3}. Furthermore, the initial scale radius plays a much smaller role in setting the amount of dust retained. 

Finally, in the extreme case of $\alpha=0.025$ the same considerations as in the case of $\alpha=10^{-2}$ apply but to a somewhat more dramatic extent. A notable exception occurs in the case of the binary discs with $R_\text{trunc}=25\text{ au}$, where the integrated dust-to-gas ratio is lower for discs with larger initial scale radius (see e.g., the leftmost green dot in the top-right-hand panel in Fig.~\ref{fig.4.1}). We will provide an explanation of this outlier as well as the dependence of $M_\text{dust}/M_\text{gas}$ on viscosity in Section~\ref{sec.5}.

\paragraph*{Slope of the dust depletion curve} Let us take into account the slope of the dust depletion curve in the upper panel of Fig.~\ref{fig.4.1}. While for $\alpha=10^{-4}\text{ and }10^{-3}$ the slope is $\sim1.5$, in the case of $\alpha=0.025$ it decreases to $\sim0.5$. As a consequence, the dust-to-gas ratio falls more quickly in the drift-dominated regime than in the purely fragmentation-dominated one, suggesting that the slope of the curve could be a way to understand which process limits the maximum grain size. Finally, when $\alpha=10^{-2}$ the slope peaks to $\sim2$ as the dust-to-gas ratio curve connects large discs in the fragmentation-dominated regime and small drift-dominated discs. \\*

To sum up, the smaller the disc tidal truncation radius the most severe the effect of binarity on disc evolution, leading ultimately to a faster removal of dust grains and dissipation of the disc. Moreover, after $t=1\text{ Myr}$ the integrated dust-to-gas ratio in binary discs increases with viscosity and the initial disc scale radius.

\paragraph*{Dependence on time} The lower panel in Fig.~\ref{fig.4.1} shows the dependence of the integrated dust-to-gas ratio on the disc tidal truncation radius, for different values of viscosity and the initial disc scale radius, after $t=3\text{ Myr}$. The triangles identify cases where the integrated dust-to-gas ratio is very small, specifically $M_\text{dust}/M_\text{gas} < 10^{-6}$.

\begin{figure*}
    \centering
	\includegraphics[width=\textwidth]{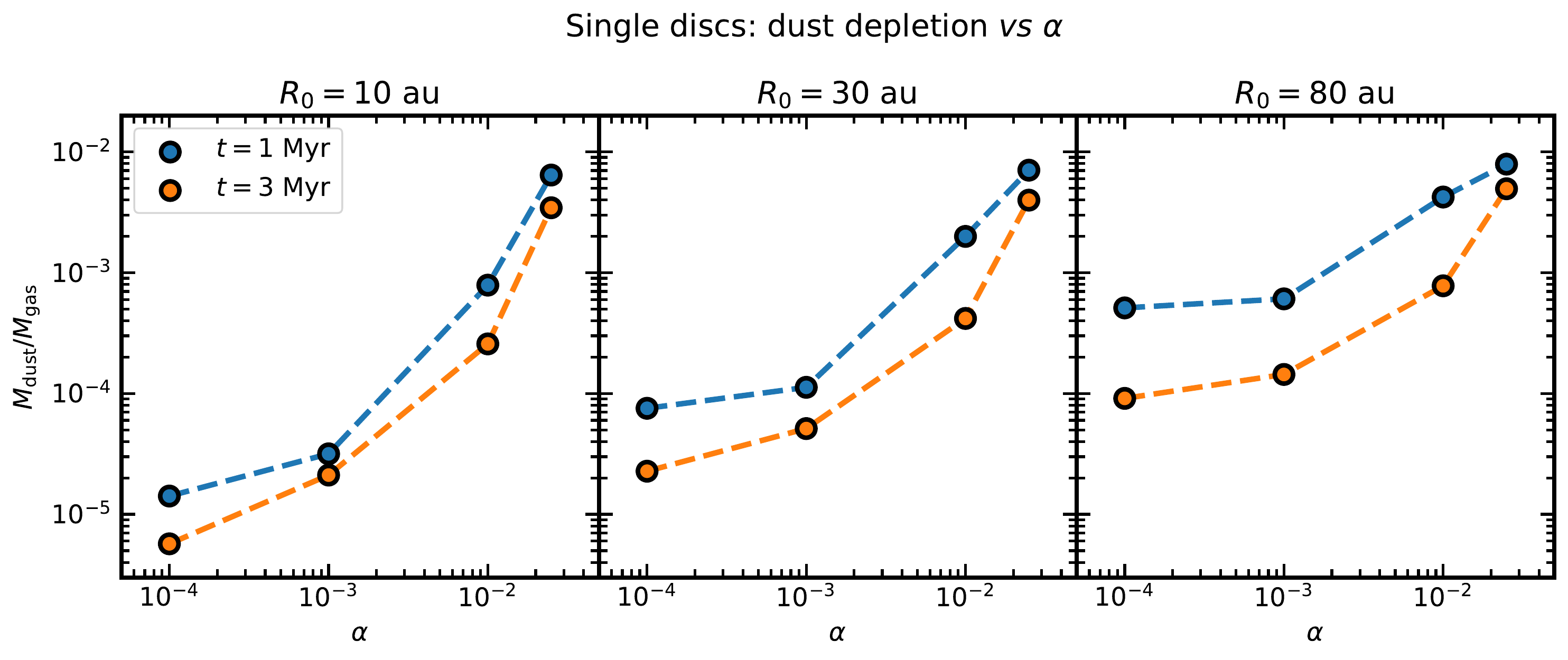}
    \caption{The integrated dust-to-gas ratio, $M_\text{dust}/M_\text{gas}$, is plotted as a function of $\alpha$, for different values of the initial disc scale radius, $R_0$, after $t=1\text{ Myr}$ and $t=3\text{ Myr}$, in the case of a single-star disc.}
    \label{fig.5.1}
\end{figure*}

Let us first focus on the $\alpha=10^{-4}$ and $\alpha=10^{-3}$ case. Confronting discs with the same initial disc scale radius in the upper and lower panels of Fig.~\ref{fig.4.1} it can be noticed that after $t=3\text{ Myr}$ the integrated dust-to-gas ratio decreased by a factor of $\sim10$, but the other features (dependence on $R_\text{trunc}$ and $R_0$, saturation radius...) are exactly the same. This suggests that in these cases the binary viscous time scales are long enough that even after $t=3\text{ Myr}$ those discs have not evolved much in the gas (i.e., the surface density is close to the initial value). Moving to larger viscosities the previous pattern changes. For $\alpha=10^{-2}$, a significant decrease in $t_\text{boundary}$, due to the increased viscosity, explains the absence of saturation which is evident from Fig.~\ref{fig.4.1}. Apparently, as in this case $t_\text{boundary}\leq 3\text{ Myr}$ regardless of $R_0$, all discs expanded enough to reach the outer boundary and the evolution of binary discs proves to be considerably different from that of single-star discs. Analogous considerations on saturation apply in the case of $\alpha=0.025$. A look at the slope of the curves, which is now $\sim1.2-1.3$ for every $\alpha$, suggests that after $t=3\text{ Myr}$ dust removal is essentially determined by radial drift. 

Remarkably, while up to $t=1\text{ Myr}$ discs with a high $\alpha$ retain more dust, after $t=3\text{ Myr}$ we see the emergence of a region of the parameter space where the opposite behaviour happens. A number of discs have been completely dispersed in the dust at high $\alpha$ (as the triangles at small $R_\text{trunc}$ in the lower panel of Fig.~\ref{fig.4.1} suggest), but not for lower viscosities. This means that, differently from single-star discs, $M_\text{dust}/M_\text{gas}$ is not a monotonically increasing function of $\alpha$. 

All in all, studying how the integrated dust-to-gas ratio changes after $t=3\text{ Myr}$ suggests that the differences between single and binary discs due to tidal effects increase with time.

\section{Dependence of the dust-to-gas ratio on viscosity: a closer look}\label{sec.5}
In this Section we investigate more in detail the dependence of the dust-to-gas ratio on disc viscosity in binaries in order to answer those questions previously left open: why does $M_\text{dust}/M_\text{gas}$ not always increase with $\alpha$ and why does changing $R_0$ affect less the dust-to-gas ratio for larger viscosities?

\paragraph*{The single-star disc case} Fig.~\ref{fig.5.1} shows the dependence of the integrated dust-to-gas ratio on $\alpha$, for different values of the initial disc scale radius, after $t=1\text{ Myr}$ and $t=3\text{ Myr}$, in the case of a single-star disc. Clearly $M_\text{dust}/M_\text{gas}$ monotonically increases with $\alpha$. This behaviour can be explained as follows. For small values of $\alpha$, the disc is in the drift-dominated regime already after $t=1\text{ Myr}$ and a significant amount of dust has been removed by radial drift, reducing the dust-to-gas ratio. As $\alpha$ increases, larger regions of the inner disc are in the fragmentation-dominated regime thus reducing the radial drift efficiency. Moreover, viscous expansion becomes more and more important, and the small dust grains entrained with the gas in the outer disc are transported at larger radii, where both the growth and drift time scales are long enough to prevent a fast depletion of the solids. In the case of extremely high values of $\alpha$, the entire disc is fragmentation-dominated and the dust is removed only when the gas is accreted. It then follows that the integrated dust-to-gas ratio is roughly equivalent to its initial value. As we already highlighted, the dust is progressively better coupled with the gas for higher viscosities. The better coupled the dust is, not only the higher the dust-to-gas ratio at the end of the simulation, but also the smaller the dependence on $R_0$ because the dust tends to be accreted with the gas.

\begin{figure*}
    \centering
    \includegraphics[width=0.999\textwidth]{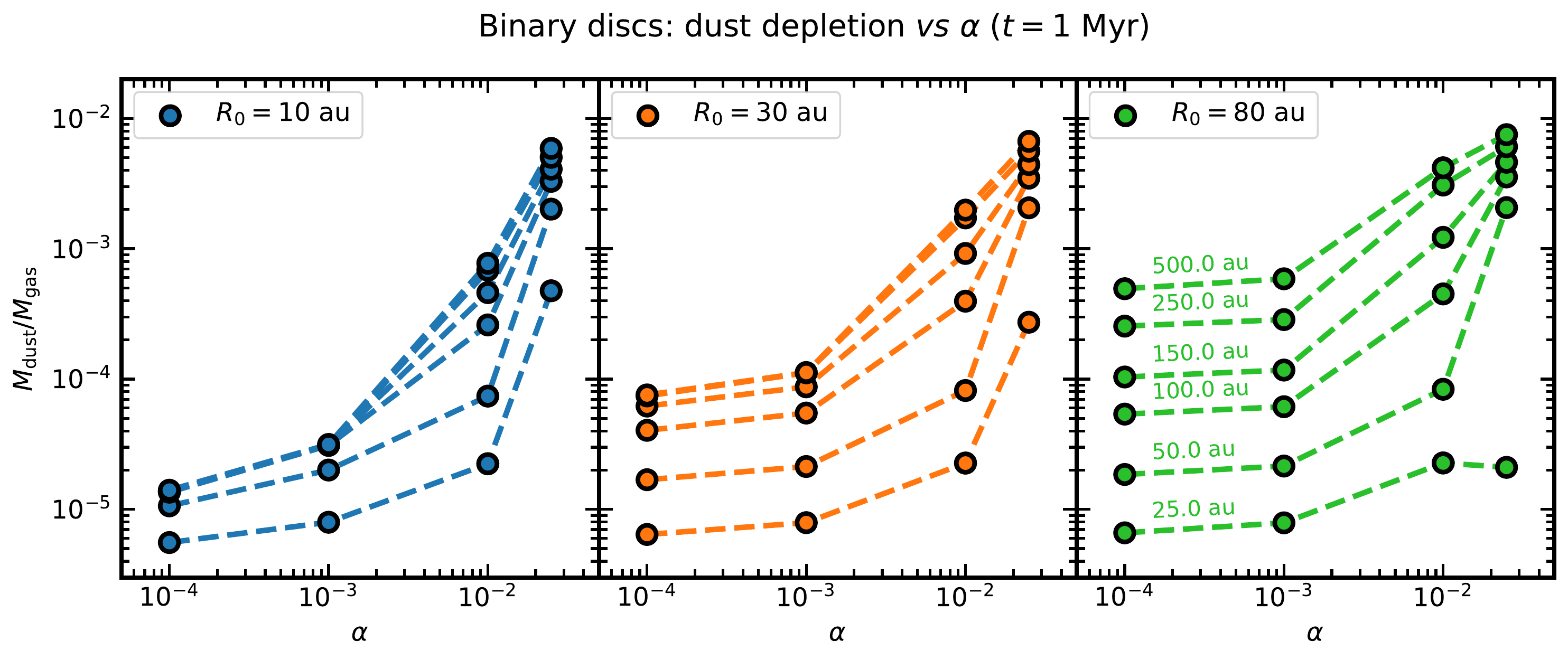}
    \includegraphics[width=0.999\textwidth]{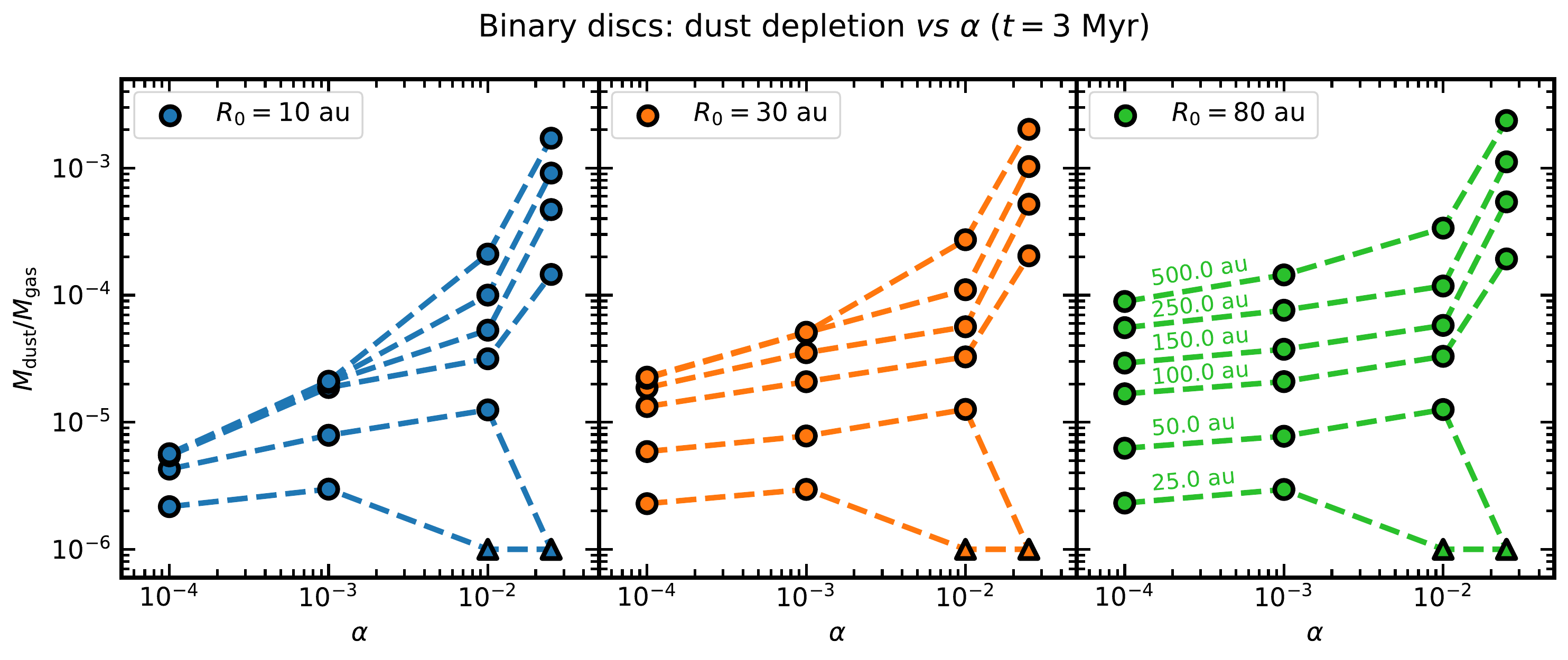}
    \caption{\textbf{Upper panel:} The integrated dust-to-gas ratio, $M_\text{dust}/M_\text{gas}$, is plotted as a function of $\alpha$ for different values of the tidal truncation radius, $R_\text{trunc}$, and the initial disc scale radius, $R_0$, after $t=1\text{ Myr}$. The dashed lines connect discs with the same tidal truncation radius, which is outlined in the rightmost panel. \textbf{Lower panel:} Same as in the upper panel but after $t=3\text{ Myr}$. The triangles identify upper limits for the integrated dust-to-gas ratio, once a sensitivity threshold has been set to $M_\text{dust}/M_\text{gas}=10^{-6}$.}
    \label{fig.5.2}
\end{figure*}

\paragraph*{The binary disc case} The upper panel in Fig.~\ref{fig.5.2} shows the dependence of the integrated dust-to-gas ratio on $\alpha$, for different values of $R_\text{trunc}$ and $R_0$, after $t=1\text{ Myr}$. Each broken-dashed line connects binary discs with the same tidal truncation radius, which is recorded in the rightmost figure. The integrated dust-to-gas ratio monotonically increases with $\alpha$, as in the case of single-star discs, regardless of the tidal truncation radius and the initial scale radius. Moreover, those discs with larger $R_\text{trunc}$ attain the same values of $M_\text{dust}/M_\text{gas}$ as single-star discs do (see Fig.~\ref{fig.5.1} for a comparison), further supporting the evidence of saturation of the dust-to-gas ratio in large binaries already outlined in Fig.~\ref{fig.4.1}.

An exception to this tendency can be noticed: in the rightmost panel of Fig.~\ref{fig.5.2}, in the disc with $R_\text{trunc}=25\text{ au}$ the integrated dust-to-gas ratio does not increase with $\alpha$. Because of the presence of a closed outer boundary, the time scale of gas and dust evolution in this disc is significantly shorter than in the single-star one with the same parameters, bringing about a faster depletion of dust grains. For this reason, after just $t=1\text{ Myr}$ a large part of the disc is in the drift-dominated regime and even monomer-size grains noticeably move radially. The reason why this effect takes place only for the smallest and most viscous disc in our sample is due to the fact that its viscous time scale is the shortest among all the other binary discs. Consequently, after just $t=1\text{ Myr}$ the disc has been substantially depleted of dust. Moreover, as discs with a larger scale radius are less massive in the inner regions, the effect of binarity is more pronounced in the case of $R_0=80\text{ au}$ because its initial dust surface density is lower than for $R_0=10\text{ and }30\text{ au}$. Consequently larger regions of the disc become drift-dominated earlier and the disc disperses faster.

\begin{figure}
    \centering
    \includegraphics[width=\columnwidth]{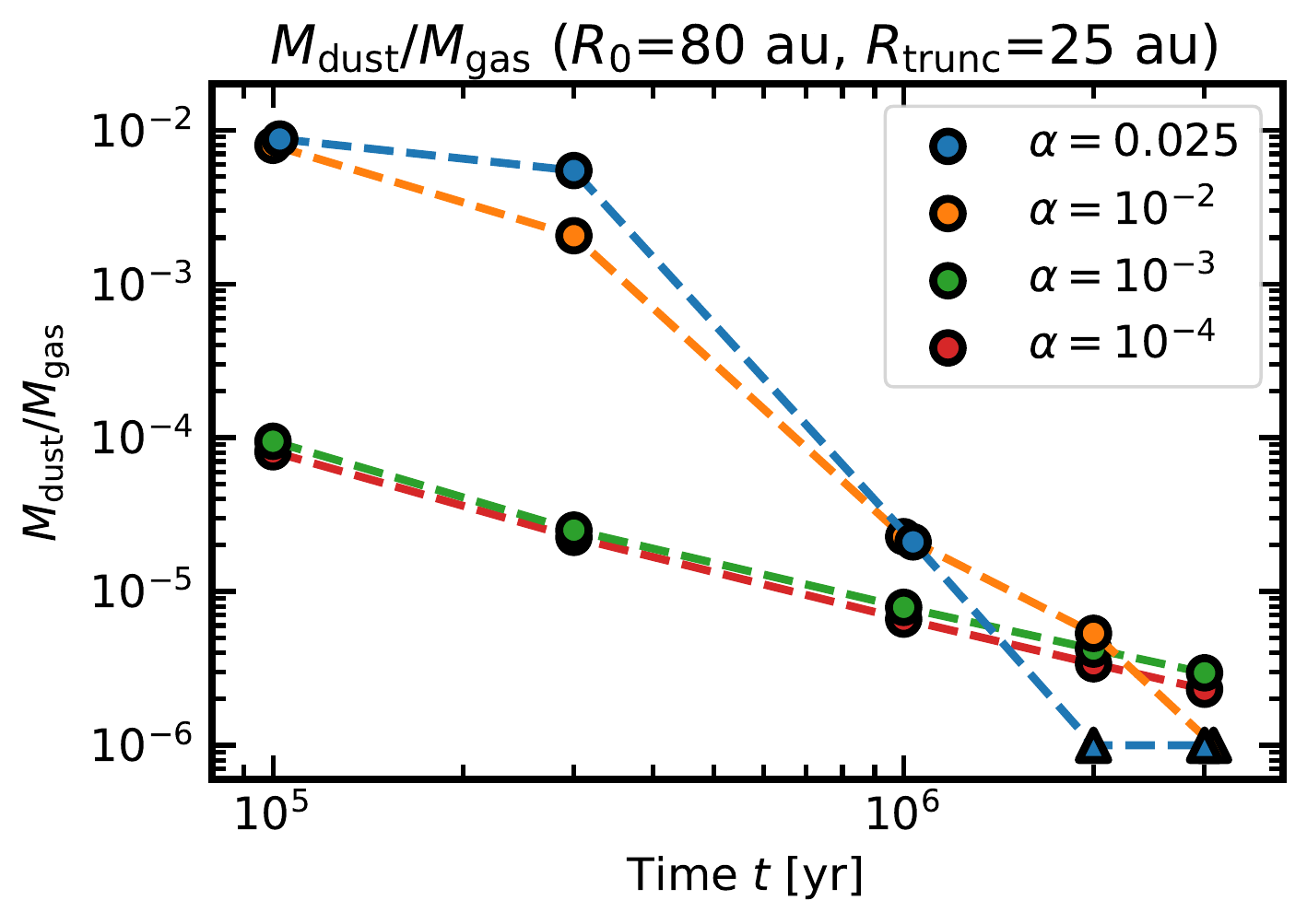}
    \caption{Time-dependence of the integrated dust-to-gas ratio in the binary disc with $R_\text{trunc}=25\text{ au}$, $R_0=80\text{ au}$ and different values of $\alpha$. The triangles identify upper limits for the integrated dust-to-gas ratio, once a sensitivity threshold has been set to $M_\text{dust}/M_\text{gas}=10^{-6}$. In the case of overlapping points, the blue series has been slightly shifted from its right time-position.}
    \label{fig.5.3}
\end{figure}

In Fig.~\ref{fig.5.3} we prove the previous argument to be true, showing the time-dependence of the integrated dust-to-gas ratio in the binary discs with $R_\text{trunc}=25\text{ au}$, $R_0=80\text{ au}$ and different values of $\alpha$. The triangles identify upper limits for the integrated dust-to-gas ratio, once a sensitivity threshold has been set to $M_\text{dust}/M_\text{gas}=10^{-6}$. As can be seen, while at low viscosities the integrated dust-to-gas ratio falls as a power law with time, for higher values of $\alpha$, and in particular for $\alpha=0.025$, it remains steady for the first dynamical time scales of disc evolution and then plummets below the threshold.

Finally, as in the case of single-star discs it is possible to see from Fig.~\ref{fig.5.2} that $R_0$ has a larger effect on the integrated dust-to-gas ratio for lower viscosities. As for freely expanding discs dust grains are more shielded from radial drift in the outer regions of extended discs at lower viscosities.

The lower panel in Fig.~\ref{fig.5.2} displays the same correlation as in the upper panel but after $t=3\text{ Myr}$. The triangles identify where the integrated dust-to-gas ratio $M_\text{dust}/M_\text{gas}<10^{-6}$. In large, less viscous binary discs $M_\text{dust}/M_\text{gas}$ monotonically increases with $\alpha$ as in the single-star case and saturation occurs in the widest binaries. In smaller binaries, instead, similar considerations as for the $t=1\text{ Myr}$ outlier apply are are now visible not only in the $R_\text{trunc}=25\text{ au}$ case but also in larger discs. Because of a fast reduction of the dust surface density, practically all the disc is drift-dominated and dust grains are removed by radial drift, which acts dispersing the disc and reducing the integrated dust-to-gas ratio to negligible values. 

To conclude, the effect of binarity on dusty discs is that of sharply shortening their lifetime. The higher the viscosity, the longer the disc survives. However, if the binary is tight enough, $R_\text{trunc}\lesssim30\text{ au}$, the trend reverses and discs with high viscosity experience a faster depletion.

\section{Implications on planet formation}\label{sec.6}

While in wide multiple systems the presence of a companion has a very limited impact on planet formation, it has been also largely recognised that both the fraction of disc-bearing and planet-hosting stars in binaries decrease sharply with stellar separation. For example, \citet{Kraus+12_2012ApJ...745...19K} show that two thirds of the discs in binaries with projected separation $a_\text{p}<40\text{ au}$ disperse faster than a million years. In addition, \citet{Kraus+16_2016AJ....152....8K} find a reduction in the fraction of multiple stars with projected separation $a_\text{p}<100\text{ au}$ hosting exoplanets. For these reasons, the presence of massive planets in binary systems with small $a_\text{p}\sim20\text{ au}$, such as HD~196885 \citep{Correia+08_2008A&A...479..271C}, HD~41004 \citep{Zucker+04_2004A&A...426..695Z}, $\gamma$~Cephei \citep{Hatzes+03_2003ApJ...599.1383H}, Gliese~86 \citep{Queloz+00_2000A&A...354...99Q}, HD~7449 \citep{Dumesque+11_2011A&A...535A..55D}, HD~87646 \citep{Ma+16_2016AJ....152..112M} and HD~176051 \citep{Muterspaugh+10_2010AJ....140.1657M} is puzzling (for further considerations see also the reviews of \citealt{Thebault&Haghighipour15_2015pes..book..309T} and \citealt{Marzari&Thebault20_2019Galax...7...84M}).

Following a complementary approach, several works have focused on the properties of exoplanets in binaries (e.g., \citealt{Moe&Kratter19_2019arXiv191201699M,Bonavita&Desidera20_2020Galax...8...16B,Hirsch+20_2020arXiv201209190H}), proving that close-in companions suppress planet formation, while planet frequency is similar in wide binaries and singles. However stellar multiplicity does not always have a negative effect: for example, \citet{Desidera&Barbieri07_2007A&A...462..345D} show that massive planets with short orbital period are more common in tight binaries.

What our models suggest is that the relative lifetime of discs around single stars and in binary systems is strongly dependent on the disc viscosity, $\alpha$, and the tidal truncation radius, $R_\text{trunc}$. In particular, highly viscous discs with $R_\text{trunc}\geq50\text{ au}$ retain as much dust as single-star discs do for the first Myr. However, as time goes on, their short viscous time scale determines a fast dispersal. On the contrary, in proto-planetary discs with a low viscosity, radial drift removes dust very quickly: after a million years discs with $R_\text{trunc}\leq100\text{ au}$ have almost completely lost their dust. However, these discs will still survive a long time after the initial phase of depletion due to their large viscous time scales. In principle both those opposing scenarios could be compatible with the formation of rocky planets according to the pebble accretion model (e.g., \citealt{Johansen&Lambrechts17_2017AREPS..45..359J,Ormel17_2017ASSL..445..197O}), provided that planetesimals form early, within $\sim1\text{ Myr}$, which could be possible (e.g., according to the streaming instability mechanism, \citealt{Youdin&Goodman05_2005ApJ...620..459Y}), and that $R_\text{trunc}\geq50-100\text{ au}$, otherwise implausibly fast planetary embryo formation would be required, in particular for the smallest and less viscous discs.

From the observational point of view, the low-viscosity scenario seems to be favoured. Indeed, \citet{Barenfeld+19_2019ApJ...878...45B} find that the fraction of close companions in systems with discs in Taurus and Upper Scorpius are comparable, suggesting that the stellar multiplicity has no substantial effect on the disc dispersal after $t\sim1-2\text{ Myr}$. This is consistent with our results for $\alpha=10^{-4}\text{ and }10^{-3}$ discs (see e.g., Fig.~\ref{fig.4.1} and the relative discussion), where the bulk of the dust is accreted before $t\sim1\text{ Myr}$ and no considerable modifications to the integrated dust-to-gas ratio are found at later times. Yet it should be mentioned that the values of $M_\mathrm{dust}$ in our models are well below the observational expectations. On the other hand, it is also possible that highly viscous binaries exist. However, due to their very fast dust-depletion time scale, discs in close multiple systems ($R_\text{trunc}\lesssim25\text{ au}$) should be undetectable after just a million years. Indeed, when small and relatively long-living discs are observed, such as in the case of the ET Cha system \citep{Ginski+20_2020arXiv200705274G}, a strongly accreting $\sim5-8$ Myr old T Tauri star with a close companion at a separation of roughly 10 au, low values of the disc viscosity are generally advised (e.g., $\alpha=10^{-4}$ in \citealt{Ginski+20_2020arXiv200705274G}).

The results of several numerical models of disc evolution in binary systems show that unless viscosity is small planet formation in binaries is strongly inhibited: e.g., \citet{Muller&Kley12_2012A&A...539A..18M} and \citet{Jang-Condell15_2015ApJ...799..147J} suggest\footnote{\citet{Jang-Condell15_2015ApJ...799..147J} results are also compatible with higher viscosities but only if implausibly high mass accretion rates are considered.} that $\alpha\lesssim10^{-3}$ should be considered for planet formation to be viable in binaries. Nevertheless, a comparison with the previously quoted works is difficult due to the different initial conditions and assumptions.

This picture significantly changes if we take into account any mechanisms inhibiting the radial flow of dust throughout the disc (e.g., \citealt{Pinilla+12_2012A&A...538A.114P}), such as the formation of rings, gaps or dust traps or even the pile-up of dust at the water snowline (e.g., \citealt{Drazkowska+16_2016A&A...594A.105D}). In this case dust is retained for longer and planets might form later during the disc lifetime, insofar as the dust traps are not created by planets but by some other mechanisms. Moreover, the dust traps could even be sites of planetesimal formation through the streaming instability \citep{Youdin&Goodman05_2005ApJ...620..459Y}, further supporting the possibility of planet formation in binaries.

The general outcome of the works that investigated circumstellar binary disc evolution and planet formation so far is that the presence of a stellar companion makes the latter process more difficult. This is in qualitative agreement with our results. \citet{Jang-Condell15_2015ApJ...799..147J} suggests that planet formation in binaries should be viable for a wider range of binary separations (and eccentricities) than ours. However, their models are in steady-state (no time evolution of the gas surface density is considered) and take into account a given composition and a fixed dust size thought the disc to infer the dust mass. In this work we showed that dust growth and migration are remarkable physical processes as far as the disc evolution is concerned: neglecting those effects leads to a substantial \textit{over}estimation of the disc solid content.

As far as gas giants are concerned, the formation of more massive rocky cores requires more dust and then either earlier planetesimal formation or larger truncation radii. In addition, it is required that their hosting discs live long enough ($t\gtrsim5-7\text{ Myr}$) for the proto-planets to accrete relevant quantities of gas, which is possible only if discs do not disperse too fast. In our models this is generally true for the less viscous discs, whose gas accretion time scales are the longest, but fails in the case of small, highly viscous discs, which disperse too fast also in the gas. Although in this paper disc evolution has been studied only for $t=3\text{ Myr}$, our models provide useful lower limits on the binary separation for discs to retain enough gas to form Jupiter-like planets. For example, in the $\alpha=10^{-3}$ case, only the shortest discs, with $R_\text{trunc}\leq10\text{ au}$, are not massive enough, with a total gas mass budget of $M_\text{gas}\leq2\times10^{-3}M_\odot\sim2M_\text{Jup}$ already after $t=3\text{ Myr}$. Remarkably, for $\alpha=10^{-2}$ this happens already for discs with $R_\text{trunc}\leq100\text{ au}$. Despite this estimate being unrefined, it is clear that gas accretion is easily achieved in either low-viscosity or wide binaries.

To conclude, our models show that the closest binaries do not retain dust grains long enough to form planets according to the classical core-accretion model and that larger truncation radii and fast embryo formation are required. In fact, the enhanced effectiveness of radial drift in binaries could be beneficial for planet formation in the pebble-accretion scenario if planetesimals are assembled fast enough or if dust traps were formed in the early epochs of the disc life. In both cases, as expected, the study of proto-planetary disc evolution in binary systems supports the hypothesis that planet formation has to occur early in the disc lifetime (e.g., \citealt{Tychoniec+20_2020A&A...640A..19T}).

\subsection{Model limitations and comparison with previous studies}
Binary disc evolution and planet formation in multiple stellar systems are difficult to study because of the high number of variables in those processes. Exploring several dynamical configurations comes with the necessity of limiting e.g., the physical dimensions taken into account. This is why we decided to adopt simplified one-dimensional models, instead of relying on higher-dimensional hydrodynamic codes. In fact, several binarity effects are difficult if not impossible to consider in one-dimensional models, such as the development of disc and orbital eccentricities, binary precession, disc warps and spirals, tidal heating, material exchange between circumstellar discs as well as between circumbinary and circumstellar discs... Those effects themselves depend on the orbital parameters of the binary systems such as separation, inclination and initial eccentricity and can only be studied properly with two- or three-dimensional simulations (e.g., \citealt{Muller&Kley12_2012A&A...539A..18M,Picogna&Marzari13_2013A&A...556A.148P}).

It is difficult to assess without dedicated simulations how such effects influence our model results, with particular reference to dust evolution in binaries (possibly halting dust settling, drift and grain collisional velocities, opposing coagulation, e.g., \citealt{Picogna&Marzari13_2013A&A...556A.148P}). Moreover, those simulations being computationally expensive, it would be prohibitive to evolve the system on secular timescales, in particular if dust is included. What is expected is that eccentric streamlines would favour higher velocity collisions of solids and easier fragmentation. However, even though eccentricity would favour a higher effective \textit{gas} viscosity (e.g., due to a faster angular momentum dissipation in spirals), it is not straightforward to assess if the same higher viscosity is responsible for \textit{dust} evolution, fragmentation (eq.~\ref{eq.2.3}), diffusion and settling.

\citet{Nelson00_2000ApJ...537L..65N} and \citet{Picogna&Marzari13_2013A&A...556A.148P} suggested that in close binaries viscous heating can be at the origin of high disc temperatures, which in turn could lead to vaporisation of water, making grains less sticky and challenging dust coagulation. This effect is not completely negative to us: dry grains, deprived of water ice, would fragment more easily because the threshold fragmentation velocity for silicates is lower ($u_\text{f}=1\text{ m s}^{-1}$, e.g., \citealt{Blum&Wurm08_2008ARA&A..46...21B}). Therefore, discs would be fragmentation dominated up to larger radii, preventing substantial grain loss by radial drift. Potentially more dust could be retained for longer periods favouring planet formation in binaries closer than previously thought. However, disc viscous heating depends on the disc accretion rate that decreases on secular time scales, leading to disc cooling (with respect to the \citealt{Nelson00_2000ApJ...537L..65N} and \citealt{Picogna&Marzari13_2013A&A...556A.148P} models). Then, potentially more favourable conditions for dust coagulation in discs can be obtained

\section{Conclusions}\label{sec.7}
In this paper we took into account the secular evolution of gas and dust in planet-forming discs around each component of a binary system, under the hypothesis that their evolution is independent of each other. The effects of the stellar multiplicity on the disc evolution were modelled imposing a zero-flux closed-outer-boundary condition on gas and dust dynamics. As for the dust, we employed the same models of grain growth and radial drift already enforced in similar studies dealing with single-star discs \citep{Rosotti+19b_2019MNRAS.486L..63R,Rosotti+19a_2019MNRAS.486.4829R}. Here are our main results:
\begin{itemize}
    \item regardless of the initial parameters, the presence of a stellar companion affects the evolution of planet-forming discs in binary systems. Nonetheless, the effects of binarity on gas and dust evolution depend on the value of the tidal truncation radius: wider binaries are less influenced by the presence of a companion and their evolution resembles that of single-star discs;
    \item as pointed out in previous works (e.g., \citealt{Rosotti&Clarke18_2018MNRAS.473.5630R}), gas evolution is faster in binaries than in single-star discs;
    \item dust evolution is affected by the presence of stellar companion, too. In particular, radial drift is a faster more effective mechanism of dust removal in binaries: the shorter the tidal truncation radius the faster dust is depleted. After just $\sim1\text{ Myr}$, for typical disc parameters such as $R_0=10\text{ au}$ and $\alpha=10^{-3}$, in a binary disc with $R_\text{trunc}=10\text{ au}$ the integrated dust-to-gas ratio is roughly $M_\text{dust}/M_\text{gas}\sim10^{-6}$, a factor of 10 smaller than in a single-star disc with the same initial parameters and becomes practically undetectable after $3\text{ Myr}$; 
    \item in the case of close binaries, the integrated dust-to-gas ratio is not always a monotonically increasing function of $\alpha$ as it is for single-star discs.
\end{itemize}

Finally, we took into account the implications of our results on planet formation. In particular, in none our closest binary models a relevant quantity of dust is retained long enough for rocky planets and gas giant cores to form according to the classical core-accretion model. Those discs prove to be significantly dust depleted already after $t\lesssim1\text{ Myr}$. This suggests that planet-formation undergoes a significant inhibition in multiple stellar systems and the only chance for this process to be effective is to happen on a very short time scale. However, if dust traps formed early enough, the enhancement of radial drift could be beneficial for planet formation in binaries both promoting planetesimal formation through the streaming instability and fast growth by pebble drift.

\section*{Acknowledgements}
We thank the anonymous referee for their helpful comments. F.Z. is grateful to Cathie Clarke and the IoA group for insightful discussions. He acknowledges support from the Erasmus+ Traineeship program and IUSS for his MSc thesis internship in Leiden as well as a Science and Technology Facilities Council (STFC) studentship and the Cambridge European Scholarship. G.R. acknowledges support from the Netherlands Organisation for Scientific Research (NWO, program number 016.Veni.192.233) and from an STFC Ernest Rutherford Fellowship (grant number ST/T003855/1). This project has received funding from the European Union’s Horizon 2020 research and innovation programme under the Marie Sklodowska-Curie grant agreement No 823823 (Dustbusters RISE project). Software: \texttt{numpy} \citep{numpy20_2020Natur.585..357H}, \texttt{matplotlib} \citep{matplotlib_Hunter:2007}, \texttt{scipy} \citep{scipy_2020SciPy-NMeth}, \texttt{JupyterNotebook} \citep{Jupyter}.

\section*{Data Availability}

The code used in this paper is publicly available on GitHub at \texttt{github.com/rbooth200/DiscEvolution}.



\bibliographystyle{mnras}
\bibliography{references}

\begin{thebibliography}{}
\makeatletter
\relax
\def\mn@urlcharsother{\let\do\@makeother \do\$\do\&\do\#\do\^\do\_\do\%\do\~}
\def\mn@doi{\begingroup\mn@urlcharsother \@ifnextchar [ {\mn@doi@}
  {\mn@doi@[]}}
\def\mn@doi@[#1]#2{\def\@tempa{#1}\ifx\@tempa\@empty \href
  {http://dx.doi.org/#2} {doi:#2}\else \href {http://dx.doi.org/#2} {#1}\fi
  \endgroup}
\def\mn@eprint#1#2{\mn@eprint@#1:#2::\@nil}
\def\mn@eprint@arXiv#1{\href {http://arxiv.org/abs/#1} {{\tt arXiv:#1}}}
\def\mn@eprint@dblp#1{\href {http://dblp.uni-trier.de/rec/bibtex/#1.xml}
  {dblp:#1}}
\def\mn@eprint@#1:#2:#3:#4\@nil{\def\@tempa {#1}\def\@tempb {#2}\def\@tempc
  {#3}\ifx \@tempc \@empty \let \@tempc \@tempb \let \@tempb \@tempa \fi \ifx
  \@tempb \@empty \def\@tempb {arXiv}\fi \@ifundefined
  {mn@eprint@\@tempb}{\@tempb:\@tempc}{\expandafter \expandafter \csname
  mn@eprint@\@tempb\endcsname \expandafter{\@tempc}}}

\bibitem[\protect\citeauthoryear{{Akeson} \& {Jensen}}{{Akeson} \&
  {Jensen}}{2014}]{Akeson&Jensen14_2014ApJ...784...62A}
{Akeson} R.~L.,  {Jensen} E.~L.~N.,  2014, \mn@doi [\apj]
  {10.1088/0004-637X/784/1/62}, \href
  {https://ui.adsabs.harvard.edu/abs/2014ApJ...784...62A} {784, 62}

\bibitem[\protect\citeauthoryear{{Akeson}, {Jensen}, {Carpenter}, {Ricci},
  {Laos}, {Nogueira}  \& {Suen-Lewis}}{{Akeson}
  et~al.}{2019}]{Akeson+19_2019ApJ...872..158A}
{Akeson} R.~L.,  {Jensen} E. L.~N.,  {Carpenter} J.,  {Ricci} L.,  {Laos} S.,
  {Nogueira} N.~F.,   {Suen-Lewis} E.~M.,  2019, \mn@doi [\apj]
  {10.3847/1538-4357/aaff6a}, \href
  {https://ui.adsabs.harvard.edu/abs/2019ApJ...872..158A} {872, 158}

\bibitem[\protect\citeauthoryear{{Alexander}}{{Alexander}}{2012}]{Alexander12_2012ApJ...757L..29A}
{Alexander} R.,  2012, \mn@doi [\apjl] {10.1088/2041-8205/757/2/L29}, \href
  {https://ui.adsabs.harvard.edu/abs/2012ApJ...757L..29A} {757, L29}

\bibitem[\protect\citeauthoryear{{Aly} \& {Lodato}}{{Aly} \&
  {Lodato}}{2020}]{Aly&Lodato20_2020MNRAS.492.3306A}
{Aly} H.,  {Lodato} G.,  2020, \mn@doi [\mnras] {10.1093/mnras/stz3633}, \href
  {https://ui.adsabs.harvard.edu/abs/2020MNRAS.492.3306A} {492, 3306}

\bibitem[\protect\citeauthoryear{{Andrews}}{{Andrews}}{2020}]{Andrews20_2020arXiv200105007A}
{Andrews} S.~M.,  2020, arXiv e-prints, \href
  {https://ui.adsabs.harvard.edu/abs/2020arXiv200105007A} {p. arXiv:2001.05007}

\bibitem[\protect\citeauthoryear{{Ansdell} et~al.,}{{Ansdell}
  et~al.}{2018}]{Ansdell+18_2018ApJ...859...21A}
{Ansdell} M.,  et~al., 2018, \mn@doi [\apj] {10.3847/1538-4357/aab890}, \href
  {https://ui.adsabs.harvard.edu/abs/2018ApJ...859...21A} {859, 21}

\bibitem[\protect\citeauthoryear{{Armitage} \& {Natarajan}}{{Armitage} \&
  {Natarajan}}{2002}]{Armitage&Natarajan02_2002ApJ...567L...9A}
{Armitage} P.~J.,  {Natarajan} P.,  2002, \mn@doi [\apjl] {10.1086/339770},
  \href {https://ui.adsabs.harvard.edu/abs/2002ApJ...567L...9A} {567, L9}

\bibitem[\protect\citeauthoryear{{Artymowicz} \& {Lubow}}{{Artymowicz} \&
  {Lubow}}{1994}]{Artymowicz&Lubow94_1994ApJ...421..651A}
{Artymowicz} P.,  {Lubow} S.~H.,  1994, \mn@doi [\apj] {10.1086/173679}, \href
  {https://ui.adsabs.harvard.edu/abs/1994ApJ...421..651A} {421, 651}

\bibitem[\protect\citeauthoryear{{Barenfeld} et~al.,}{{Barenfeld}
  et~al.}{2019}]{Barenfeld+19_2019ApJ...878...45B}
{Barenfeld} S.~A.,  et~al., 2019, \mn@doi [\apj] {10.3847/1538-4357/ab1e50},
  \href {https://ui.adsabs.harvard.edu/abs/2019ApJ...878...45B} {878, 45}

\bibitem[\protect\citeauthoryear{{Bath} \& {Pringle}}{{Bath} \&
  {Pringle}}{1981}]{Bath&Pringle81_1981MNRAS.194..967B}
{Bath} G.~T.,  {Pringle} J.~E.,  1981, \mn@doi [\mnras]
  {10.1093/mnras/194.4.967}, \href
  {https://ui.adsabs.harvard.edu/abs/1981MNRAS.194..967B} {194, 967}

\bibitem[\protect\citeauthoryear{{Birnstiel} \& {Andrews}}{{Birnstiel} \&
  {Andrews}}{2014}]{Birnstiel&Andrews+14_2014ApJ...780..153B}
{Birnstiel} T.,  {Andrews} S.~M.,  2014, \mn@doi [\apj]
  {10.1088/0004-637X/780/2/153}, \href
  {https://ui.adsabs.harvard.edu/abs/2014ApJ...780..153B} {780, 153}

\bibitem[\protect\citeauthoryear{{Birnstiel}, {Dullemond}  \&
  {Brauer}}{{Birnstiel} et~al.}{2009}]{Birnstiel+09_2009A&A...503L...5B}
{Birnstiel} T.,  {Dullemond} C.~P.,   {Brauer} F.,  2009, \mn@doi [\aap]
  {10.1051/0004-6361/200912452}, \href
  {https://ui.adsabs.harvard.edu/abs/2009A&A...503L...5B} {503, L5}

\bibitem[\protect\citeauthoryear{{Birnstiel}, {Dullemond}  \&
  {Brauer}}{{Birnstiel} et~al.}{2010}]{Birnstiel+10_2010A&A...513A..79B}
{Birnstiel} T.,  {Dullemond} C.~P.,   {Brauer} F.,  2010, \mn@doi [\aap]
  {10.1051/0004-6361/200913731}, \href
  {https://ui.adsabs.harvard.edu/abs/2010A&A...513A..79B} {513, A79}

\bibitem[\protect\citeauthoryear{{Birnstiel}, {Klahr}  \&
  {Ercolano}}{{Birnstiel} et~al.}{2012}]{Birnstiel+12_2012A&A...539A.148B}
{Birnstiel} T.,  {Klahr} H.,   {Ercolano} B.,  2012, \mn@doi [\aap]
  {10.1051/0004-6361/201118136}, \href
  {https://ui.adsabs.harvard.edu/abs/2012A&A...539A.148B} {539, A148}

\bibitem[\protect\citeauthoryear{{Blum} \& {Wurm}}{{Blum} \&
  {Wurm}}{2008}]{Blum&Wurm08_2008ARA&A..46...21B}
{Blum} J.,  {Wurm} G.,  2008, \mn@doi [\araa]
  {10.1146/annurev.astro.46.060407.145152}, \href
  {https://ui.adsabs.harvard.edu/abs/2008ARA&A..46...21B} {46, 21}

\bibitem[\protect\citeauthoryear{{Bonavita} \& {Desidera}}{{Bonavita} \&
  {Desidera}}{2020}]{Bonavita&Desidera20_2020Galax...8...16B}
{Bonavita} M.,  {Desidera} S.,  2020, \mn@doi [Galaxies]
  {10.3390/galaxies8010016}, \href
  {https://ui.adsabs.harvard.edu/abs/2020Galax...8...16B} {8, 16}

\bibitem[\protect\citeauthoryear{{Booth} \& {Clarke}}{{Booth} \&
  {Clarke}}{2021}]{Booth&Clarke21_2021MNRAS.502.1569B}
{Booth} R.~A.,  {Clarke} C.~J.,  2021, \mn@doi [\mnras]
  {10.1093/mnras/stab090}, \href
  {https://ui.adsabs.harvard.edu/abs/2021MNRAS.502.1569B} {502, 1569}

\bibitem[\protect\citeauthoryear{{Booth}, {Clarke}, {Madhusudhan}  \&
  {Ilee}}{{Booth} et~al.}{2017}]{Booth+17_2017MNRAS.469.3994B}
{Booth} R.~A.,  {Clarke} C.~J.,  {Madhusudhan} N.,   {Ilee} J.~D.,  2017,
  \mn@doi [\mnras] {10.1093/mnras/stx1103}, \href
  {https://ui.adsabs.harvard.edu/abs/2017MNRAS.469.3994B} {469, 3994}

\bibitem[\protect\citeauthoryear{{Brauer}, {Dullemond}  \& {Henning}}{{Brauer}
  et~al.}{2008}]{Brauer+08_2008A&A...480..859B}
{Brauer} F.,  {Dullemond} C.~P.,   {Henning} T.,  2008, \mn@doi [\aap]
  {10.1051/0004-6361:20077759}, \href
  {https://ui.adsabs.harvard.edu/abs/2008A&A...480..859B} {480, 859}

\bibitem[\protect\citeauthoryear{{Chachan}, {Booth}, {Triaud}  \&
  {Clarke}}{{Chachan} et~al.}{2019}]{Chachan+19_2019MNRAS.489.3896C}
{Chachan} Y.,  {Booth} R.~A.,  {Triaud} A. H.~M.~J.,   {Clarke} C.,  2019,
  \mn@doi [\mnras] {10.1093/mnras/stz2404}, \href
  {https://ui.adsabs.harvard.edu/abs/2019MNRAS.489.3896C} {489, 3896}

\bibitem[\protect\citeauthoryear{{Chiang} \& {Goldreich}}{{Chiang} \&
  {Goldreich}}{1997}]{Chiang&Goldreich97_1997ApJ...490..368C}
{Chiang} E.~I.,  {Goldreich} P.,  1997, \mn@doi [\apj] {10.1086/304869}, \href
  {https://ui.adsabs.harvard.edu/abs/1997ApJ...490..368C} {490, 368}

\bibitem[\protect\citeauthoryear{{Cieza} et~al.,}{{Cieza}
  et~al.}{2009}]{Cieza+09_2009ApJ...696L..84C}
{Cieza} L.~A.,  et~al., 2009, \mn@doi [\apjl] {10.1088/0004-637X/696/1/L84},
  \href {https://ui.adsabs.harvard.edu/abs/2009ApJ...696L..84C} {696, L84}

\bibitem[\protect\citeauthoryear{{Correia} et~al.,}{{Correia}
  et~al.}{2008}]{Correia+08_2008A&A...479..271C}
{Correia} A.~C.~M.,  et~al., 2008, \mn@doi [\aap] {10.1051/0004-6361:20078908},
  \href {https://ui.adsabs.harvard.edu/abs/2008A&A...479..271C} {479, 271}

\bibitem[\protect\citeauthoryear{{Cox} et~al.,}{{Cox}
  et~al.}{2017}]{Cox+17_2017ApJ...851...83C}
{Cox} E.~G.,  et~al., 2017, \mn@doi [\apj] {10.3847/1538-4357/aa97e2}, \href
  {https://ui.adsabs.harvard.edu/abs/2017ApJ...851...83C} {851, 83}

\bibitem[\protect\citeauthoryear{{Daemgen}, {Correia}  \&
  {Petr-Gotzens}}{{Daemgen} et~al.}{2012}]{Daemgen+12_2012A&A...540A..46D}
{Daemgen} S.,  {Correia} S.,   {Petr-Gotzens} M.~G.,  2012, \mn@doi [\aap]
  {10.1051/0004-6361/201118314}, \href
  {https://ui.adsabs.harvard.edu/abs/2012A&A...540A..46D} {540, A46}

\bibitem[\protect\citeauthoryear{{Daemgen}, {Petr-Gotzens}, {Correia},
  {Teixeira}, {Brandner}, {Kley}  \& {Zinnecker}}{{Daemgen}
  et~al.}{2013}]{Daemgen+13_2013A&A...554A..43D}
{Daemgen} S.,  {Petr-Gotzens} M.~G.,  {Correia} S.,  {Teixeira} P.~S.,
  {Brandner} W.,  {Kley} W.,   {Zinnecker} H.,  2013, \mn@doi [\aap]
  {10.1051/0004-6361/201321220}, \href
  {https://ui.adsabs.harvard.edu/abs/2013A&A...554A..43D} {554, A43}

\bibitem[\protect\citeauthoryear{{Desidera} \& {Barbieri}}{{Desidera} \&
  {Barbieri}}{2007}]{Desidera&Barbieri07_2007A&A...462..345D}
{Desidera} S.,  {Barbieri} M.,  2007, \mn@doi [\aap]
  {10.1051/0004-6361:20066319}, \href
  {https://ui.adsabs.harvard.edu/abs/2007A&A...462..345D} {462, 345}

\bibitem[\protect\citeauthoryear{{Dipierro}, {Laibe}, {Alexander}  \&
  {Hutchison}}{{Dipierro} et~al.}{2018}]{Dipierro+18_2018MNRAS.479.4187D}
{Dipierro} G.,  {Laibe} G.,  {Alexander} R.,   {Hutchison} M.,  2018, \mn@doi
  [\mnras] {10.1093/mnras/sty1701}, \href
  {https://ui.adsabs.harvard.edu/abs/2018MNRAS.479.4187D} {479, 4187}

\bibitem[\protect\citeauthoryear{{Doyle} et~al.,}{{Doyle}
  et~al.}{2011}]{Doyle+11_2011Sci...333.1602D}
{Doyle} L.~R.,  et~al., 2011, \mn@doi [Science] {10.1126/science.1210923},
  \href {https://ui.adsabs.harvard.edu/abs/2011Sci...333.1602D} {333, 1602}

\bibitem[\protect\citeauthoryear{{Dr{\k{a}}{\.z}kowska}, {Alibert}  \&
  {Moore}}{{Dr{\k{a}}{\.z}kowska}
  et~al.}{2016}]{Drazkowska+16_2016A&A...594A.105D}
{Dr{\k{a}}{\.z}kowska} J.,  {Alibert} Y.,   {Moore} B.,  2016, \mn@doi [\aap]
  {10.1051/0004-6361/201628983}, \href
  {https://ui.adsabs.harvard.edu/abs/2016A&A...594A.105D} {594, A105}

\bibitem[\protect\citeauthoryear{{Dumusque} et~al.,}{{Dumusque}
  et~al.}{2011}]{Dumesque+11_2011A&A...535A..55D}
{Dumusque} X.,  et~al., 2011, \mn@doi [\aap] {10.1051/0004-6361/201117148},
  \href {https://ui.adsabs.harvard.edu/abs/2011A&A...535A..55D} {535, A55}

\bibitem[\protect\citeauthoryear{{Dumusque} et~al.,}{{Dumusque}
  et~al.}{2012}]{Dumesque+12_2012Natur.491..207D}
{Dumusque} X.,  et~al., 2012, \mn@doi [\nat] {10.1038/nature11572}, \href
  {https://ui.adsabs.harvard.edu/abs/2012Natur.491..207D} {491, 207}

\bibitem[\protect\citeauthoryear{{Ercolano} \& {Pascucci}}{{Ercolano} \&
  {Pascucci}}{2017}]{Ercolano&Pascucci17_2017RSOS....470114E}
{Ercolano} B.,  {Pascucci} I.,  2017, \mn@doi [Royal Society Open Science]
  {10.1098/rsos.170114}, \href
  {https://ui.adsabs.harvard.edu/abs/2017RSOS....470114E} {4, 170114}

\bibitem[\protect\citeauthoryear{{Facchini} et~al.,}{{Facchini}
  et~al.}{2019}]{Facchini+19_2019A&A...626L...2F}
{Facchini} S.,  et~al., 2019, \mn@doi [\aap] {10.1051/0004-6361/201935496},
  \href {https://ui.adsabs.harvard.edu/abs/2019A&A...626L...2F} {626, L2}

\bibitem[\protect\citeauthoryear{{Flaherty} et~al.,}{{Flaherty}
  et~al.}{2020}]{Flaherty+20_2020ApJ...895..109F}
{Flaherty} K.,  et~al., 2020, \mn@doi [\apj] {10.3847/1538-4357/ab8cc5}, \href
  {https://ui.adsabs.harvard.edu/abs/2020ApJ...895..109F} {895, 109}

\bibitem[\protect\citeauthoryear{{Fragner}, {Nelson}  \& {Kley}}{{Fragner}
  et~al.}{2011}]{Fragner+11_2011A&A...528A..40F}
{Fragner} M.~M.,  {Nelson} R.~P.,   {Kley} W.,  2011, \mn@doi [\aap]
  {10.1051/0004-6361/201015378}, \href
  {https://ui.adsabs.harvard.edu/abs/2011A&A...528A..40F} {528, A40}

\bibitem[\protect\citeauthoryear{{G{\'a}rate}, {Birnstiel},
  {Dr{\k{a}}{\.z}kowska}  \& {Stammler}}{{G{\'a}rate}
  et~al.}{2020}]{Garate+20_2020A&A...635A.149G}
{G{\'a}rate} M.,  {Birnstiel} T.,  {Dr{\k{a}}{\.z}kowska} J.,   {Stammler}
  S.~M.,  2020, \mn@doi [\aap] {10.1051/0004-6361/201936067}, \href
  {https://ui.adsabs.harvard.edu/abs/2020A&A...635A.149G} {635, A149}

\bibitem[\protect\citeauthoryear{{Ginski} et~al.,}{{Ginski}
  et~al.}{2020}]{Ginski+20_2020arXiv200705274G}
{Ginski} C.,  et~al., 2020, arXiv e-prints, \href
  {https://ui.adsabs.harvard.edu/abs/2020arXiv200705274G} {p. arXiv:2007.05274}

\bibitem[\protect\citeauthoryear{{Goldreich} \& {Tremaine}}{{Goldreich} \&
  {Tremaine}}{1979}]{Goldreich&Tremaine79_1979ApJ...233..857G}
{Goldreich} P.,  {Tremaine} S.,  1979, \mn@doi [\apj] {10.1086/157448}, \href
  {https://ui.adsabs.harvard.edu/abs/1979ApJ...233..857G} {233, 857}

\bibitem[\protect\citeauthoryear{{Goldreich} \& {Tremaine}}{{Goldreich} \&
  {Tremaine}}{1980}]{Goldreich&Tremaine80_1980ApJ...241..425G}
{Goldreich} P.,  {Tremaine} S.,  1980, \mn@doi [\apj] {10.1086/158356}, \href
  {https://ui.adsabs.harvard.edu/abs/1980ApJ...241..425G} {241, 425}

\bibitem[\protect\citeauthoryear{{Gundlach} \& {Blum}}{{Gundlach} \&
  {Blum}}{2015}]{Gundlach&Blum15_2015ApJ...798...34G}
{Gundlach} B.,  {Blum} J.,  2015, \mn@doi [\apj] {10.1088/0004-637X/798/1/34},
  \href {https://ui.adsabs.harvard.edu/abs/2015ApJ...798...34G} {798, 34}

\bibitem[\protect\citeauthoryear{{Haisch}, {Lada}  \& {Lada}}{{Haisch}
  et~al.}{2001}]{Haisch+01_2001ApJ...553L.153H}
{Haisch} Karl~E. J.,  {Lada} E.~A.,   {Lada} C.~J.,  2001, \mn@doi [\apjl]
  {10.1086/320685}, \href
  {https://ui.adsabs.harvard.edu/abs/2001ApJ...553L.153H} {553, L153}

\bibitem[\protect\citeauthoryear{{Harris}, {Andrews}, {Wilner}  \&
  {Kraus}}{{Harris} et~al.}{2012}]{Harris+12_2012ApJ...751..115H}
{Harris} R.~J.,  {Andrews} S.~M.,  {Wilner} D.~J.,   {Kraus} A.~L.,  2012,
  \mn@doi [\apj] {10.1088/0004-637X/751/2/115}, \href
  {https://ui.adsabs.harvard.edu/abs/2012ApJ...751..115H} {751, 115}

\bibitem[\protect\citeauthoryear{{Harris} et~al.,}{{Harris}
  et~al.}{2020}]{numpy20_2020Natur.585..357H}
{Harris} C.~R.,  et~al., 2020, \mn@doi [\nat] {10.1038/s41586-020-2649-2},
  \href {https://ui.adsabs.harvard.edu/abs/2020Natur.585..357H} {585, 357}

\bibitem[\protect\citeauthoryear{{Hatzes}}{{Hatzes}}{2016}]{Hatzes16_2016SSRv..205..267H}
{Hatzes} A.~P.,  2016, \mn@doi [\ssr] {10.1007/s11214-016-0246-3}, \href
  {https://ui.adsabs.harvard.edu/abs/2016SSRv..205..267H} {205, 267}

\bibitem[\protect\citeauthoryear{{Hatzes}, {Cochran}, {Endl}, {McArthur},
  {Paulson}, {Walker}, {Campbell}  \& {Yang}}{{Hatzes}
  et~al.}{2003}]{Hatzes+03_2003ApJ...599.1383H}
{Hatzes} A.~P.,  {Cochran} W.~D.,  {Endl} M.,  {McArthur} B.,  {Paulson} D.~B.,
   {Walker} G. A.~H.,  {Campbell} B.,   {Yang} S.,  2003, \mn@doi [\apj]
  {10.1086/379281}, \href
  {https://ui.adsabs.harvard.edu/abs/2003ApJ...599.1383H} {599, 1383}

\bibitem[\protect\citeauthoryear{{Hirsch} et~al.,}{{Hirsch}
  et~al.}{2020}]{Hirsch+20_2020arXiv201209190H}
{Hirsch} L.~A.,  et~al., 2020, arXiv e-prints, \href
  {https://ui.adsabs.harvard.edu/abs/2020arXiv201209190H} {p. arXiv:2012.09190}

\bibitem[\protect\citeauthoryear{Hunter}{Hunter}{2007}]{matplotlib_Hunter:2007}
Hunter J.~D.,  2007, \mn@doi [Computing in Science \& Engineering]
  {10.1109/MCSE.2007.55}, 9, 90

\bibitem[\protect\citeauthoryear{{Jang-Condell}}{{Jang-Condell}}{2015}]{Jang-Condell15_2015ApJ...799..147J}
{Jang-Condell} H.,  2015, \mn@doi [\apj] {10.1088/0004-637X/799/2/147}, \href
  {https://ui.adsabs.harvard.edu/abs/2015ApJ...799..147J} {799, 147}

\bibitem[\protect\citeauthoryear{{Jang-Condell}, {Mugrauer}  \&
  {Schmidt}}{{Jang-Condell} et~al.}{2008}]{Jang-Condell+08_2008ApJ...683L.191J}
{Jang-Condell} H.,  {Mugrauer} M.,   {Schmidt} T.,  2008, \mn@doi [\apjl]
  {10.1086/591791}, \href
  {https://ui.adsabs.harvard.edu/abs/2008ApJ...683L.191J} {683, L191}

\bibitem[\protect\citeauthoryear{{Jensen}, {Mathieu}  \& {Fuller}}{{Jensen}
  et~al.}{1994}]{Jensen+94_1994ApJ...429L..29J}
{Jensen} E. L.~N.,  {Mathieu} R.~D.,   {Fuller} G.~A.,  1994, \mn@doi [\apjl]
  {10.1086/187405}, \href
  {https://ui.adsabs.harvard.edu/abs/1994ApJ...429L..29J} {429, L29}

\bibitem[\protect\citeauthoryear{{Jensen}, {Mathieu}  \& {Fuller}}{{Jensen}
  et~al.}{1996}]{Jensen+96_1996ApJ...458..312J}
{Jensen} E. L.~N.,  {Mathieu} R.~D.,   {Fuller} G.~A.,  1996, \mn@doi [\apj]
  {10.1086/176814}, \href
  {https://ui.adsabs.harvard.edu/abs/1996ApJ...458..312J} {458, 312}

\bibitem[\protect\citeauthoryear{{Johansen} \& {Lambrechts}}{{Johansen} \&
  {Lambrechts}}{2017}]{Johansen&Lambrechts17_2017AREPS..45..359J}
{Johansen} A.,  {Lambrechts} M.,  2017, \mn@doi [Annual Review of Earth and
  Planetary Sciences] {10.1146/annurev-earth-063016-020226}, \href
  {https://ui.adsabs.harvard.edu/abs/2017AREPS..45..359J} {45, 359}

\bibitem[\protect\citeauthoryear{Kluyver et~al.,}{Kluyver
  et~al.}{2016}]{Jupyter}
Kluyver T.,  et~al., 2016, in Loizides F.,  Scmidt B.,  eds, Positioning and
  Power in Academic Publishing: Players, Agents and Agendas. IOS Press, pp
  87--90, \url {https://eprints.soton.ac.uk/403913/}

\bibitem[\protect\citeauthoryear{{Kraus}, {Ireland}, {Hillenbrand}  \&
  {Martinache}}{{Kraus} et~al.}{2012}]{Kraus+12_2012ApJ...745...19K}
{Kraus} A.~L.,  {Ireland} M.~J.,  {Hillenbrand} L.~A.,   {Martinache} F.,
  2012, \mn@doi [\apj] {10.1088/0004-637X/745/1/19}, \href
  {https://ui.adsabs.harvard.edu/abs/2012ApJ...745...19K} {745, 19}

\bibitem[\protect\citeauthoryear{{Kraus}, {Ireland}, {Huber}, {Mann}  \&
  {Dupuy}}{{Kraus} et~al.}{2016}]{Kraus+16_2016AJ....152....8K}
{Kraus} A.~L.,  {Ireland} M.~J.,  {Huber} D.,  {Mann} A.~W.,   {Dupuy} T.~J.,
  2016, \mn@doi [\aj] {10.3847/0004-6256/152/1/8}, \href
  {https://ui.adsabs.harvard.edu/abs/2016AJ....152....8K} {152, 8}

\bibitem[\protect\citeauthoryear{{Laibe} \& {Price}}{{Laibe} \&
  {Price}}{2014}]{Laibe&Price14_2014MNRAS.444.1940L}
{Laibe} G.,  {Price} D.~J.,  2014, \mn@doi [\mnras] {10.1093/mnras/stu1367},
  \href {https://ui.adsabs.harvard.edu/abs/2014MNRAS.444.1940L} {444, 1940}

\bibitem[\protect\citeauthoryear{{Lin} \& {Papaloizou}}{{Lin} \&
  {Papaloizou}}{1979}]{Lin&Papaloizou79_1979MNRAS.186..799L}
{Lin} D.~N.~C.,  {Papaloizou} J.,  1979, \mn@doi [\mnras]
  {10.1093/mnras/186.4.799}, \href
  {https://ui.adsabs.harvard.edu/abs/1979MNRAS.186..799L} {186, 799}

\bibitem[\protect\citeauthoryear{{Lin} \& {Papaloizou}}{{Lin} \&
  {Papaloizou}}{1986}]{Lin&Papaloizou86_1986ApJ...309..846L}
{Lin} D.~N.~C.,  {Papaloizou} J.,  1986, \mn@doi [\apj] {10.1086/164653}, \href
  {https://ui.adsabs.harvard.edu/abs/1986ApJ...309..846L} {309, 846}

\bibitem[\protect\citeauthoryear{{Lines}, {Leinhardt}, {Baruteau},
  {Paardekooper}  \& {Carter}}{{Lines}
  et~al.}{2015}]{Lines+15_2015A&A...582A...5L}
{Lines} S.,  {Leinhardt} Z.~M.,  {Baruteau} C.,  {Paardekooper} S.~J.,
  {Carter} P.~J.,  2015, \mn@doi [\aap] {10.1051/0004-6361/201526295}, \href
  {https://ui.adsabs.harvard.edu/abs/2015A&A...582A...5L} {582, A5}

\bibitem[\protect\citeauthoryear{{Lines}, {Leinhardt}, {Baruteau},
  {Paardekooper}  \& {Carter}}{{Lines}
  et~al.}{2016}]{Lines+16_2016A&A...590A..62L}
{Lines} S.,  {Leinhardt} Z.~M.,  {Baruteau} C.,  {Paardekooper} S.~J.,
  {Carter} P.~J.,  2016, \mn@doi [\aap] {10.1051/0004-6361/201628349}, \href
  {https://ui.adsabs.harvard.edu/abs/2016A&A...590A..62L} {590, A62}

\bibitem[\protect\citeauthoryear{{Lodato} \& {Clarke}}{{Lodato} \&
  {Clarke}}{2004}]{Lodato&Clarke04_2004MNRAS.353..841L}
{Lodato} G.,  {Clarke} C.~J.,  2004, \mn@doi [\mnras]
  {10.1111/j.1365-2966.2004.08112.x}, \href
  {https://ui.adsabs.harvard.edu/abs/2004MNRAS.353..841L} {353, 841}

\bibitem[\protect\citeauthoryear{{Lodato}, {Nayakshin}, {King}  \&
  {Pringle}}{{Lodato} et~al.}{2009}]{Lodato+09_2009MNRAS.398.1392L}
{Lodato} G.,  {Nayakshin} S.,  {King} A.~R.,   {Pringle} J.~E.,  2009, \mn@doi
  [\mnras] {10.1111/j.1365-2966.2009.15179.x}, \href
  {https://ui.adsabs.harvard.edu/abs/2009MNRAS.398.1392L} {398, 1392}

\bibitem[\protect\citeauthoryear{{Lodato}, {Scardoni}, {Manara}  \&
  {Testi}}{{Lodato} et~al.}{2017}]{Lodato+17_2017MNRAS.472.4700L}
{Lodato} G.,  {Scardoni} C.~E.,  {Manara} C.~F.,   {Testi} L.,  2017, \mn@doi
  [\mnras] {10.1093/mnras/stx2273}, \href
  {https://ui.adsabs.harvard.edu/abs/2017MNRAS.472.4700L} {472, 4700}

\bibitem[\protect\citeauthoryear{{Lynden-Bell} \& {Pringle}}{{Lynden-Bell} \&
  {Pringle}}{1974}]{Lynden-Bell&Pringle74_1974MNRAS.168..603L}
{Lynden-Bell} D.,  {Pringle} J.~E.,  1974, \mn@doi [\mnras]
  {10.1093/mnras/168.3.603}, \href
  {https://ui.adsabs.harvard.edu/abs/1974MNRAS.168..603L} {168, 603}

\bibitem[\protect\citeauthoryear{{Ma} et~al.,}{{Ma}
  et~al.}{2016}]{Ma+16_2016AJ....152..112M}
{Ma} B.,  et~al., 2016, \mn@doi [\aj] {10.3847/0004-6256/152/5/112}, \href
  {https://ui.adsabs.harvard.edu/abs/2016AJ....152..112M} {152, 112}

\bibitem[\protect\citeauthoryear{{Manara} et~al.,}{{Manara}
  et~al.}{2019}]{Manara+19_2019A&A...628A..95M}
{Manara} C.~F.,  et~al., 2019, \mn@doi [\aap] {10.1051/0004-6361/201935964},
  \href {https://ui.adsabs.harvard.edu/abs/2019A&A...628A..95M} {628, A95}

\bibitem[\protect\citeauthoryear{{Martin}}{{Martin}}{2018}]{Martin18_2018haex.bookE.156M}
{Martin} D.~V.,  2018, {Populations of Planets in Multiple Star Systems}.
p.~156, \mn@doi{10.1007/978-3-319-55333-7_156}

\bibitem[\protect\citeauthoryear{{Martin} \& {Triaud}}{{Martin} \&
  {Triaud}}{2014}]{Martin&Triaud14_2014A&A...570A..91M}
{Martin} D.~V.,  {Triaud} A. H.~M.~J.,  2014, \mn@doi [\aap]
  {10.1051/0004-6361/201323112}, \href
  {https://ui.adsabs.harvard.edu/abs/2014A&A...570A..91M} {570, A91}

\bibitem[\protect\citeauthoryear{{Marzari} \& {Thebault}}{{Marzari} \&
  {Thebault}}{2019}]{Marzari&Thebault20_2019Galax...7...84M}
{Marzari} F.,  {Thebault} P.,  2019, \mn@doi [Galaxies]
  {10.3390/galaxies7040084}, \href
  {https://ui.adsabs.harvard.edu/abs/2019Galax...7...84M} {7, 84}

\bibitem[\protect\citeauthoryear{{Marzari}, {Thebault}, {Scholl}, {Picogna}  \&
  {Baruteau}}{{Marzari} et~al.}{2013}]{Marzari+13_2013A&A...553A..71M}
{Marzari} F.,  {Thebault} P.,  {Scholl} H.,  {Picogna} G.,   {Baruteau} C.,
  2013, \mn@doi [\aap] {10.1051/0004-6361/201220893}, \href
  {https://ui.adsabs.harvard.edu/abs/2013A&A...553A..71M} {553, A71}

\bibitem[\protect\citeauthoryear{{Moe} \& {Kratter}}{{Moe} \&
  {Kratter}}{2019}]{Moe&Kratter19_2019arXiv191201699M}
{Moe} M.,  {Kratter} K.~M.,  2019, arXiv e-prints, \href
  {https://ui.adsabs.harvard.edu/abs/2019arXiv191201699M} {p. arXiv:1912.01699}

\bibitem[\protect\citeauthoryear{{M{\"u}ller} \& {Kley}}{{M{\"u}ller} \&
  {Kley}}{2012}]{Muller&Kley12_2012A&A...539A..18M}
{M{\"u}ller} T.~W.~A.,  {Kley} W.,  2012, \mn@doi [\aap]
  {10.1051/0004-6361/201118202}, \href
  {https://ui.adsabs.harvard.edu/abs/2012A&A...539A..18M} {539, A18}

\bibitem[\protect\citeauthoryear{{Muterspaugh} et~al.,}{{Muterspaugh}
  et~al.}{2010}]{Muterspaugh+10_2010AJ....140.1657M}
{Muterspaugh} M.~W.,  et~al., 2010, \mn@doi [\aj]
  {10.1088/0004-6256/140/6/1657}, \href
  {https://ui.adsabs.harvard.edu/abs/2010AJ....140.1657M} {140, 1657}

\bibitem[\protect\citeauthoryear{{Nelson}}{{Nelson}}{2000}]{Nelson00_2000ApJ...537L..65N}
{Nelson} A.~F.,  2000, \mn@doi [\apjl] {10.1086/312752}, \href
  {https://ui.adsabs.harvard.edu/abs/2000ApJ...537L..65N} {537, L65}

\bibitem[\protect\citeauthoryear{{Ormel}}{{Ormel}}{2017}]{Ormel17_2017ASSL..445..197O}
{Ormel} C.~W.,  2017, {The Emerging Paradigm of Pebble Accretion}.
p.~197, \mn@doi{10.1007/978-3-319-60609-5_7}

\bibitem[\protect\citeauthoryear{{Orosz} et~al.,}{{Orosz}
  et~al.}{2019}]{Orosz+19_2019AJ....157..174O}
{Orosz} J.~A.,  et~al., 2019, \mn@doi [\aj] {10.3847/1538-3881/ab0ca0}, \href
  {https://ui.adsabs.harvard.edu/abs/2019AJ....157..174O} {157, 174}

\bibitem[\protect\citeauthoryear{{Paardekooper}, {Th{\'e}bault}  \&
  {Mellema}}{{Paardekooper} et~al.}{2008}]{Paardekooper+08_2008MNRAS.386..973P}
{Paardekooper} S.~J.,  {Th{\'e}bault} P.,   {Mellema} G.,  2008, \mn@doi
  [\mnras] {10.1111/j.1365-2966.2008.13080.x}, \href
  {https://ui.adsabs.harvard.edu/abs/2008MNRAS.386..973P} {386, 973}

\bibitem[\protect\citeauthoryear{{Paczynski}}{{Paczynski}}{1977}]{Paczynski77_1977ApJ...216..822P}
{Paczynski} B.,  1977, \mn@doi [\apj] {10.1086/155526}, \href
  {https://ui.adsabs.harvard.edu/abs/1977ApJ...216..822P} {216, 822}

\bibitem[\protect\citeauthoryear{{Pani{\'c}} et~al.,}{{Pani{\'c}}
  et~al.}{2020}]{Panic+20_2020arXiv201207901P}
{Pani{\'c}} O.,  et~al., 2020, arXiv e-prints, \href
  {https://ui.adsabs.harvard.edu/abs/2020arXiv201207901P} {p. arXiv:2012.07901}

\bibitem[\protect\citeauthoryear{{Papaloizou} \& {Pringle}}{{Papaloizou} \&
  {Pringle}}{1977}]{Papaloizou&Pringle77_1977MNRAS.181..441P}
{Papaloizou} J.,  {Pringle} J.~E.,  1977, \mn@doi [\mnras]
  {10.1093/mnras/181.3.441}, \href
  {https://ui.adsabs.harvard.edu/abs/1977MNRAS.181..441P} {181, 441}

\bibitem[\protect\citeauthoryear{{Patience}, {Akeson}  \& {Jensen}}{{Patience}
  et~al.}{2008}]{Patience+08_2008ApJ...677..616P}
{Patience} J.,  {Akeson} R.~L.,   {Jensen} E.~L.~N.,  2008, \mn@doi [\apj]
  {10.1086/526394}, \href
  {https://ui.adsabs.harvard.edu/abs/2008ApJ...677..616P} {677, 616}

\bibitem[\protect\citeauthoryear{{Pichardo}, {Sparke}  \& {Aguilar}}{{Pichardo}
  et~al.}{2005}]{Pichardo+05_2005MNRAS.359..521P}
{Pichardo} B.,  {Sparke} L.~S.,   {Aguilar} L.~A.,  2005, \mn@doi [\mnras]
  {10.1111/j.1365-2966.2005.08905.x}, \href
  {https://ui.adsabs.harvard.edu/abs/2005MNRAS.359..521P} {359, 521}

\bibitem[\protect\citeauthoryear{{Picogna} \& {Marzari}}{{Picogna} \&
  {Marzari}}{2013}]{Picogna&Marzari13_2013A&A...556A.148P}
{Picogna} G.,  {Marzari} F.,  2013, \mn@doi [\aap]
  {10.1051/0004-6361/201321860}, \href
  {https://ui.adsabs.harvard.edu/abs/2013A&A...556A.148P} {556, A148}

\bibitem[\protect\citeauthoryear{{Pinilla}, {Birnstiel}, {Ricci}, {Dullemond},
  {Uribe}, {Testi}  \& {Natta}}{{Pinilla}
  et~al.}{2012}]{Pinilla+12_2012A&A...538A.114P}
{Pinilla} P.,  {Birnstiel} T.,  {Ricci} L.,  {Dullemond} C.~P.,  {Uribe} A.~L.,
   {Testi} L.,   {Natta} A.,  2012, \mn@doi [\aap]
  {10.1051/0004-6361/201118204}, \href
  {https://ui.adsabs.harvard.edu/abs/2012A&A...538A.114P} {538, A114}

\bibitem[\protect\citeauthoryear{{Pringle}}{{Pringle}}{1981}]{Pringle81_1981ARA&A..19..137P}
{Pringle} J.~E.,  1981, \mn@doi [\araa] {10.1146/annurev.aa.19.090181.001033},
  \href {https://ui.adsabs.harvard.edu/abs/1981ARA&A..19..137P} {19, 137}

\bibitem[\protect\citeauthoryear{{Queloz} et~al.,}{{Queloz}
  et~al.}{2000}]{Queloz+00_2000A&A...354...99Q}
{Queloz} D.,  et~al., 2000, \aap, \href
  {https://ui.adsabs.harvard.edu/abs/2000A&A...354...99Q} {354, 99}

\bibitem[\protect\citeauthoryear{{Rafikov}}{{Rafikov}}{2013}]{Rafikov13_2013ApJ...774..144R}
{Rafikov} R.~R.,  2013, \mn@doi [\apj] {10.1088/0004-637X/774/2/144}, \href
  {https://ui.adsabs.harvard.edu/abs/2013ApJ...774..144R} {774, 144}

\bibitem[\protect\citeauthoryear{{Raghavan} et~al.,}{{Raghavan}
  et~al.}{2010}]{Raghavan+10_2010ApJS..190....1R}
{Raghavan} D.,  et~al., 2010, \mn@doi [\apjs] {10.1088/0067-0049/190/1/1},
  \href {https://ui.adsabs.harvard.edu/abs/2010ApJS..190....1R} {190, 1}

\bibitem[\protect\citeauthoryear{{Ribas}, {Mer{\'\i}n}, {Bouy}  \&
  {Maud}}{{Ribas} et~al.}{2014}]{Ribas+14_2014A&A...561A..54R}
{Ribas} {\'A}.,  {Mer{\'\i}n} B.,  {Bouy} H.,   {Maud} L.~T.,  2014, \mn@doi
  [\aap] {10.1051/0004-6361/201322597}, \href
  {https://ui.adsabs.harvard.edu/abs/2014A&A...561A..54R} {561, A54}

\bibitem[\protect\citeauthoryear{{Rodriguez} et~al.,}{{Rodriguez}
  et~al.}{2018}]{Rodriguez+18_2018ApJ...859..150R}
{Rodriguez} J.~E.,  et~al., 2018, \mn@doi [\apj] {10.3847/1538-4357/aac08f},
  \href {https://ui.adsabs.harvard.edu/abs/2018ApJ...859..150R} {859, 150}

\bibitem[\protect\citeauthoryear{{Rosotti} \& {Clarke}}{{Rosotti} \&
  {Clarke}}{2018}]{Rosotti&Clarke18_2018MNRAS.473.5630R}
{Rosotti} G.~P.,  {Clarke} C.~J.,  2018, \mn@doi [\mnras]
  {10.1093/mnras/stx2769}, \href
  {https://ui.adsabs.harvard.edu/abs/2018MNRAS.473.5630R} {473, 5630}

\bibitem[\protect\citeauthoryear{{Rosotti}, {Booth}, {Tazzari}, {Clarke},
  {Lodato}  \& {Testi}}{{Rosotti}
  et~al.}{2019a}]{Rosotti+19b_2019MNRAS.486L..63R}
{Rosotti} G.~P.,  {Booth} R.~A.,  {Tazzari} M.,  {Clarke} C.,  {Lodato} G.,
  {Testi} L.,  2019a, \mn@doi [\mnras] {10.1093/mnrasl/slz064}, \href
  {https://ui.adsabs.harvard.edu/abs/2019MNRAS.486L..63R} {486, L63}

\bibitem[\protect\citeauthoryear{{Rosotti}, {Tazzari}, {Booth}, {Testi},
  {Lodato}  \& {Clarke}}{{Rosotti}
  et~al.}{2019b}]{Rosotti+19a_2019MNRAS.486.4829R}
{Rosotti} G.~P.,  {Tazzari} M.,  {Booth} R.~A.,  {Testi} L.,  {Lodato} G.,
  {Clarke} C.,  2019b, \mn@doi [\mnras] {10.1093/mnras/stz1190}, \href
  {https://ui.adsabs.harvard.edu/abs/2019MNRAS.486.4829R} {486, 4829}

\bibitem[\protect\citeauthoryear{{Shakura} \& {Sunyaev}}{{Shakura} \&
  {Sunyaev}}{1973}]{Shakura&Sunyaev73_1973A&A....24..337S}
{Shakura} N.~I.,  {Sunyaev} R.~A.,  1973, \aap, \href
  {https://ui.adsabs.harvard.edu/abs/1973A&A....24..337S} {500, 33}

\bibitem[\protect\citeauthoryear{{Silsbee} \& {Rafikov}}{{Silsbee} \&
  {Rafikov}}{2015}]{Silsbee+15_2015ApJ...798...71S}
{Silsbee} K.,  {Rafikov} R.~R.,  2015, \mn@doi [\apj]
  {10.1088/0004-637X/798/2/71}, \href
  {https://ui.adsabs.harvard.edu/abs/2015ApJ...798...71S} {798, 71}

\bibitem[\protect\citeauthoryear{{Syer} \& {Clarke}}{{Syer} \&
  {Clarke}}{1995}]{Syer&Clarke95_1995MNRAS.277..758S}
{Syer} D.,  {Clarke} C.~J.,  1995, \mn@doi [\mnras] {10.1093/mnras/277.3.758},
  \href {https://ui.adsabs.harvard.edu/abs/1995MNRAS.277..758S} {277, 758}

\bibitem[\protect\citeauthoryear{{Tazzari} \& {Lodato}}{{Tazzari} \&
  {Lodato}}{2015}]{Tazzari&Lodato15_2015MNRAS.449.1118T}
{Tazzari} M.,  {Lodato} G.,  2015, \mn@doi [\mnras] {10.1093/mnras/stv352},
  \href {https://ui.adsabs.harvard.edu/abs/2015MNRAS.449.1118T} {449, 1118}

\bibitem[\protect\citeauthoryear{{Testi} et~al.,}{{Testi}
  et~al.}{2014}]{Testi+14_2014prpl.conf..339T}
{Testi} L.,  et~al., 2014, in {Beuther} H.,  {Klessen} R.~S.,  {Dullemond}
  C.~P.,   {Henning} T.,  eds, Protostars and Planets VI. p.~339 (\mn@eprint
  {arXiv} {1402.1354}), \mn@doi{10.2458/azu_uapress_9780816531240-ch015}

\bibitem[\protect\citeauthoryear{{Thebault} \& {Haghighipour}}{{Thebault} \&
  {Haghighipour}}{2015}]{Thebault&Haghighipour15_2015pes..book..309T}
{Thebault} P.,  {Haghighipour} N.,  2015, {Planet Formation in Binaries}.
pp 309--340, \mn@doi{10.1007/978-3-662-45052-9_13}

\bibitem[\protect\citeauthoryear{{Th{\'e}bault}, {Marzari}  \&
  {Scholl}}{{Th{\'e}bault} et~al.}{2006}]{Thebault+06_2006Icar..183..193T}
{Th{\'e}bault} P.,  {Marzari} F.,   {Scholl} H.,  2006, \mn@doi [\icarus]
  {10.1016/j.icarus.2006.01.022}, \href
  {https://ui.adsabs.harvard.edu/abs/2006Icar..183..193T} {183, 193}

\bibitem[\protect\citeauthoryear{{Th{\'e}bault}, {Marzari}  \&
  {Scholl}}{{Th{\'e}bault} et~al.}{2008}]{Thebault+08_2008MNRAS.388.1528T}
{Th{\'e}bault} P.,  {Marzari} F.,   {Scholl} H.,  2008, \mn@doi [\mnras]
  {10.1111/j.1365-2966.2008.13536.x}, \href
  {https://ui.adsabs.harvard.edu/abs/2008MNRAS.388.1528T} {388, 1528}

\bibitem[\protect\citeauthoryear{{Trapman}, {Rosotti}, {Bosman}, {Hogerheijde}
  \& {van Dishoeck}}{{Trapman} et~al.}{2020}]{Trapman+20_2020arXiv200511330T}
{Trapman} L.,  {Rosotti} G.,  {Bosman} A.~D.,  {Hogerheijde} M.~R.,   {van
  Dishoeck} E.~F.,  2020, arXiv e-prints, \href
  {https://ui.adsabs.harvard.edu/abs/2020arXiv200511330T} {p. arXiv:2005.11330}

\bibitem[\protect\citeauthoryear{{Tychoniec} et~al.,}{{Tychoniec}
  et~al.}{2020}]{Tychoniec+20_2020A&A...640A..19T}
{Tychoniec} {\L}.,  et~al., 2020, \mn@doi [\aap] {10.1051/0004-6361/202037851},
  \href {https://ui.adsabs.harvard.edu/abs/2020A&A...640A..19T} {640, A19}

\bibitem[\protect\citeauthoryear{{Virtanen} et~al.,}{{Virtanen}
  et~al.}{2020}]{scipy_2020SciPy-NMeth}
{Virtanen} P.,  et~al., 2020, \mn@doi [Nature Methods]
  {https://doi.org/10.1038/s41592-019-0686-2}, \href {https://rdcu.be/b08Wh}
  {17, 261}

\bibitem[\protect\citeauthoryear{{Weidenschilling}}{{Weidenschilling}}{1977}]{Weidenschilling77_1977MNRAS.180...57W}
{Weidenschilling} S.~J.,  1977, \mn@doi [\mnras] {10.1093/mnras/180.1.57},
  \href {https://ui.adsabs.harvard.edu/abs/1977MNRAS.180...57W} {180, 57}

\bibitem[\protect\citeauthoryear{{Welsh} et~al.,}{{Welsh}
  et~al.}{2012}]{Welsh+12_2012Natur.481..475W}
{Welsh} W.~F.,  et~al., 2012, \mn@doi [\nat] {10.1038/nature10768}, \href
  {https://ui.adsabs.harvard.edu/abs/2012Natur.481..475W} {481, 475}

\bibitem[\protect\citeauthoryear{{Whipple}}{{Whipple}}{1972}]{Whipple72_1972fpp..conf..211W}
{Whipple} F.~L.,  1972, in {Elvius} A.,  ed., From Plasma to Planet. p.~211

\bibitem[\protect\citeauthoryear{{Winn} \& {Fabrycky}}{{Winn} \&
  {Fabrycky}}{2015}]{Winn&Fabrycky15_2015ARA&A..53..409W}
{Winn} J.~N.,  {Fabrycky} D.~C.,  2015, \mn@doi [\araa]
  {10.1146/annurev-astro-082214-122246}, \href
  {https://ui.adsabs.harvard.edu/abs/2015ARA&A..53..409W} {53, 409}

\bibitem[\protect\citeauthoryear{{Xie}, {Payne}, {Th{\'e}bault}, {Zhou}  \&
  {Ge}}{{Xie} et~al.}{2010}]{Xie+10_2010ApJ...724.1153X}
{Xie} J.-W.,  {Payne} M.~J.,  {Th{\'e}bault} P.,  {Zhou} J.-L.,   {Ge} J.,
  2010, \mn@doi [\apj] {10.1088/0004-637X/724/2/1153}, \href
  {https://ui.adsabs.harvard.edu/abs/2010ApJ...724.1153X} {724, 1153}

\bibitem[\protect\citeauthoryear{{Youdin} \& {Goodman}}{{Youdin} \&
  {Goodman}}{2005}]{Youdin&Goodman05_2005ApJ...620..459Y}
{Youdin} A.~N.,  {Goodman} J.,  2005, \mn@doi [\apj] {10.1086/426895}, \href
  {https://ui.adsabs.harvard.edu/abs/2005ApJ...620..459Y} {620, 459}

\bibitem[\protect\citeauthoryear{{Zsom}, {S{\'a}ndor}  \& {Dullemond}}{{Zsom}
  et~al.}{2011}]{Zsom+11_2011A&A...527A..10Z}
{Zsom} A.,  {S{\'a}ndor} Z.,   {Dullemond} C.~P.,  2011, \mn@doi [\aap]
  {10.1051/0004-6361/201015434}, \href
  {https://ui.adsabs.harvard.edu/abs/2011A&A...527A..10Z} {527, A10}

\bibitem[\protect\citeauthoryear{{Zucker}, {Mazeh}, {Santos}, {Udry}  \&
  {Mayor}}{{Zucker} et~al.}{2004}]{Zucker+04_2004A&A...426..695Z}
{Zucker} S.,  {Mazeh} T.,  {Santos} N.~C.,  {Udry} S.,   {Mayor} M.,  2004,
  \mn@doi [\aap] {10.1051/0004-6361:20040384}, \href
  {https://ui.adsabs.harvard.edu/abs/2004A&A...426..695Z} {426, 695}

\bibitem[\protect\citeauthoryear{{Zurlo} et~al.,}{{Zurlo}
  et~al.}{2020a}]{Zurlo+20b_2020MNRAS.tmp.3477Z}
{Zurlo} A.,  et~al., 2020a, \mn@doi [\mnras] {10.1093/mnras/staa3674}, \href
  {https://ui.adsabs.harvard.edu/abs/2020MNRAS.tmp.3477Z} {}

\bibitem[\protect\citeauthoryear{{Zurlo} et~al.,}{{Zurlo}
  et~al.}{2020b}]{Zurlo+20_2020MNRAS.496.5089Z}
{Zurlo} A.,  et~al., 2020b, \mn@doi [\mnras] {10.1093/mnras/staa1886}, \href
  {https://ui.adsabs.harvard.edu/abs/2020MNRAS.496.5089Z} {496, 5089}

\makeatother
\end{thebibliography}



\appendix

\section{The effects of enforcing an explicit torque}\label{app:1}
In this paper we studied the effects of the tidal truncation on gas evolution in planet-forming discs in a simplified framework. Specifically, instead of solving the viscous-evolution equation including an explicit-torque term, we imposed a zero-flux condition at the outer disc radius, $R_\text{out}$. This choice is physically motivated by the very steep radial dependence of the torque, which can be considered as an infinite potential acting on a very narrow interval. Despite being an idealisation, the zero-flux condition allows for a clear definition of the disc tidal truncation radius, $R_\text{trunc}=R_\text{out}$. This in turn reduces the complexity and degeneracy as well as the number of parameters of the problem, which historically have been among the main difficulties in studying disc evolution in binary systems. 

As it will become clear in this Section, although it has often been envisaged as the correct implementation, also employing an explicit torque comes with some degrees of idealisation and approximation. In particular, the torque needs to be smoothed and its strength to be fine-tuned to reproduce both the position of the truncation radius and the width of the gap that the stellar companion opens in the disc. The latter quantities are intrinsically ill-defined in this formulation of the problem as the key parameters to be considered are now the position of the satellite and the binary mass ratio. To obtain accurate results, it is necessary to perform a systematical comparison of the one-dimensional code predictions with more complex two- or even three-dimensional hydro-dynamical simulations.

It is then evident that a self-consistent comparison between the zero-flux and the explicit-torque formulations, which is beyond the aims of this paper, is not as straightforward as could be initially thought and comes with intrinsic difficulties which need to be addressed in a dedicated work. Hereafter we discuss some of the possible variations of our results if an explicit torque is employed. To reproduce correctly the position of the gap and to fairly compare our results with the zero-flux framework, we rely on the implementation discussed in \citet{Tazzari&Lodato15_2015MNRAS.449.1118T}.

\subsection{Numerical implementation}
We follow the prescription of \citet{Lin&Papaloizou86_1986ApJ...309..846L}, which was successfully adopted in several works and in different physical contexts \citep[e.g.,][]{Syer&Clarke95_1995MNRAS.277..758S,Armitage&Natarajan02_2002ApJ...567L...9A,Lodato&Clarke04_2004MNRAS.353..841L,Lodato+09_2009MNRAS.398.1392L}. In particular, instead of Eq.~\ref{eq.2.1}, we solve:
\begin{equation}\label{eq:A1}
    \dfrac{\partial\Sigma}{\partial t}=\dfrac{3}{R}\dfrac{\partial}{\partial R}\biggl[R^{1/2}\dfrac{\partial}{\partial R}\bigl(\nu\Sigma_\text{g}R^{1/2}\bigr)\biggr]-\dfrac{1}{R}\dfrac{\partial}{\partial R}\left[\dfrac{\Lambda_\text{t}\Sigma_\text{g}}{\Omega_\text{K}}\right],
\end{equation}
where the torque exerted by the secondary on the circumprimary disc, $\Lambda_\text{t}=\Lambda_\text{t}(a,q,R)$, is given by:
\begin{equation}\label{eq:A2}
    \Lambda_\text{t}(a,q,R) =
    \begin{dcases}
    \begin{aligned}
    -fq^2\Omega_\text{K}^2R^2\left(\dfrac{R}{\Delta}\right)^4&\text{ if }R<a, \\
    fq^2\Omega_\text{K}^2R^2\left(\dfrac{a}{\Delta}\right)^4&\text{ if }R>a.
    \end{aligned}
    \end{dcases}
\end{equation}
Here $a$ is the binary separation, $q$ is the binary mass ratio, $\Delta=\lvert R-a\lvert$ and $f$ is a normalisation factor. 

To avoid any divergences in the torque at the position of the secondary, we choose \citep[e.g,][]{Tazzari&Lodato15_2015MNRAS.449.1118T}:
\begin{equation}\label{eq:A3}
    \Delta=\max\bigl(\lvert R-a\lvert, H, R_\text{Hill}\bigr),
\end{equation}
where $R_\text{Hill}=a(q/3)^{1/3}$ is the Hill radius of the secondary. This decision can be motivated as follows. First of all, due to the gas pressure, a gap smaller than the disc vertical scale height at the position of the secondary would be re-filled of gas. Moreover, gas accretion onto the secondary needs to be considered. This is done parametrically imposing that the gap is larger then the Hill radius of the secondary.

\begin{figure}
    \centering
	\includegraphics[width=\columnwidth]{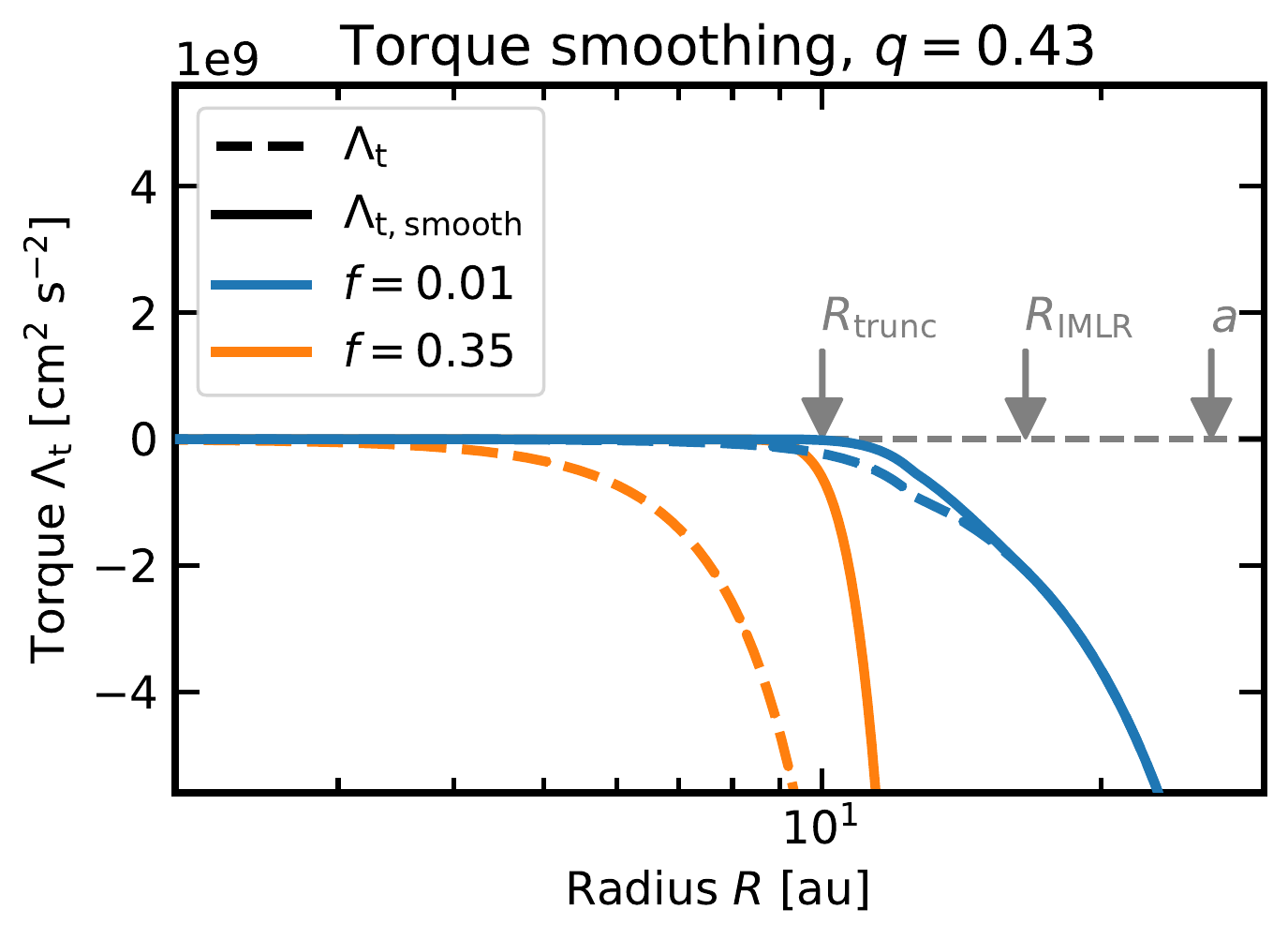}
    \caption{The tidal torque is plotted as a function of the disc radius as in Eq.~\ref{eq:A2} for $f=10^{-2}$ and $f=3.5\times10^{-1}$ (blue and orange dashed lines, respectively) in the model with $R_\text{trunc}=10\text{ au}$ and $q=0.43$. The solid lines of the same colours refer to the smoothed torque computed as in \citet{Tazzari&Lodato15_2015MNRAS.449.1118T}.}
    \label{fig.A1}
\end{figure}

As for the correct determination of the gap size and the position of the truncation radius we follow the implementation of \citet{Tazzari&Lodato15_2015MNRAS.449.1118T}. In this work the torque expression in Eq.~\ref{eq:A2} was smoothed inside the innermost and outside the outermost Lindblad resonances in order to reproduce the results of the three-dimensional SPH simulations in \citet{Artymowicz&Lubow94_1994ApJ...421..651A}. In particular, we choose $f=10^{-2}$ as in \citet{Armitage&Natarajan02_2002ApJ...567L...9A} and multiply the tidal torque by the Gaussian term:
\begin{equation}\label{eq:A4}
    \exp\left\{-\biggl(\dfrac{R-R_\mathrm{IMLR}}{\omega_\mathrm{IMLR}}\biggr)^2\right\}\text{ if }R<R_\mathrm{IMLR}
\end{equation}
where $R_\mathrm{IMLR}\sim0.59a$ is the position of the innermost Lindblad resonance and $\omega_\mathrm{IMLR}$ determines the steepness of the Gaussian term. As we are interested only in the inner gap edge we do not apply the equivalent smoothing term involving the outermost Lindblad resonance \citep[see][]{Tazzari&Lodato15_2015MNRAS.449.1118T}. $\omega_\mathrm{IMLR}$ is fine-tuned so as to have $R_\text{gap}\sim R_\text{trunc}$, which can be achieved, as shown in Fig.~\ref{fig.A1}, requiring that $\Lambda_\text{t}\sim0$ inside $R_\text{trunc}$. We obtain $\omega_\mathrm{IMLR}\sim7H$ which differs from the value reported in \citet{Tazzari&Lodato15_2015MNRAS.449.1118T} probably because of the different vertical structure of the disc and the initial condition for the gas surface density.

We solve Eq.~\ref{eq:A1} using the same finite-differences one-dimensional code employed in Section~\ref{sec.2}. Its grid, which is equally spaced in $R^{1/2}$, extends from $R_\text{in}=0.01\text{ au}$ to $R_\text{out}=10000\text{ au}$ and is made up of 1000 radial cells. If the required resolution is not achieved, which typically happens for small discs, we adopt a grid made up of 4000 cells.  
We ran a number of different models changing the position of the secondary, $a$, the binary mass ratio, $q=0.11\text{ and }0.43$, as well as the disc viscosity, $\alpha=10^{-3},\,10^{-2}$, while the initial disc scale radius was kept fixed at $R_0=10\text{ au}$. For $q=0.11$ and $q=0.43$ the inner edge of the gap is expected to be located at $R_\text{gap}\sim0.46a$ and  $R_\text{gap}\sim0.38a$, respectively \citep[see][]{Artymowicz&Lubow94_1994ApJ...421..651A,Tazzari&Lodato15_2015MNRAS.449.1118T}. In the following we will compare explicit-torque simulations with $R_\text{gap}\sim10,\,150\text{ and }500\text{ au}$ with zero-flux binary models with $R_\text{trunc}=R_\text{gap}$. Our initial condition is set as in Eq.~\ref{eq.2.6}, imposing that $\Sigma(R>R_\text{trunc})=0$.

\subsection{Gas and dust evolution: confronting the two formulations}
This Section is dedicated to a systematical comparison of the zero-flux and explicit-torque model prescriptions. 

\begin{figure*}
    \centering
	\includegraphics[width=\textwidth]{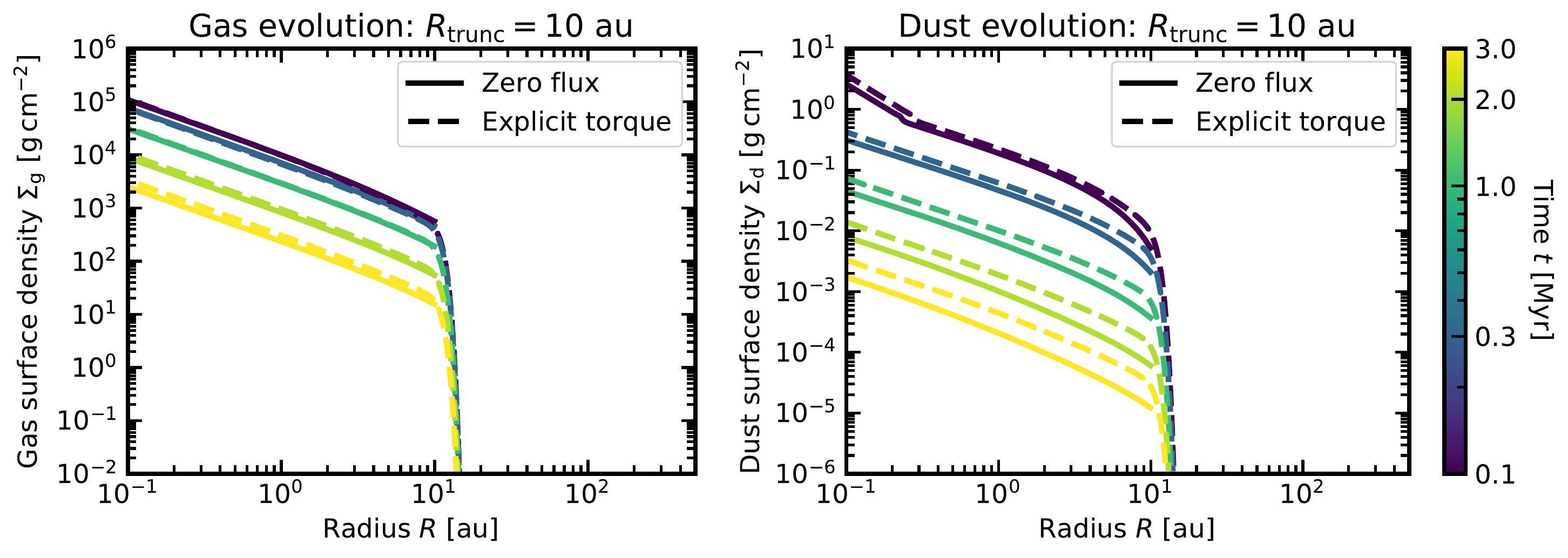}
    \caption{\textbf{Left panel:} Radial dependence of the disc gas surface density after $t=0.1,\,0.3,\,1,\,2\text{ and }3\text{ Myr}$. The solid lines identify the model enforcing the zero-flux boundary condition with $R_0=10\text{ au}$, $R_\text{trunc}=10\text{ au}$ and $\alpha=10^{-3}$. The explicit-torque solution with $R_0=10\text{ au}$, $q=0.11$, $R_\text{gap}=a/0.46\sim10\text{ au}$ and $\alpha=10^{-3}$ is plotted as a dashed line. \textbf{Right panel:} Same as in the upper panel for the disc dust surface density radial profile.}
    \label{fig.A2}
\end{figure*}

\begin{figure*}
    \centering
	\includegraphics[width=\textwidth]{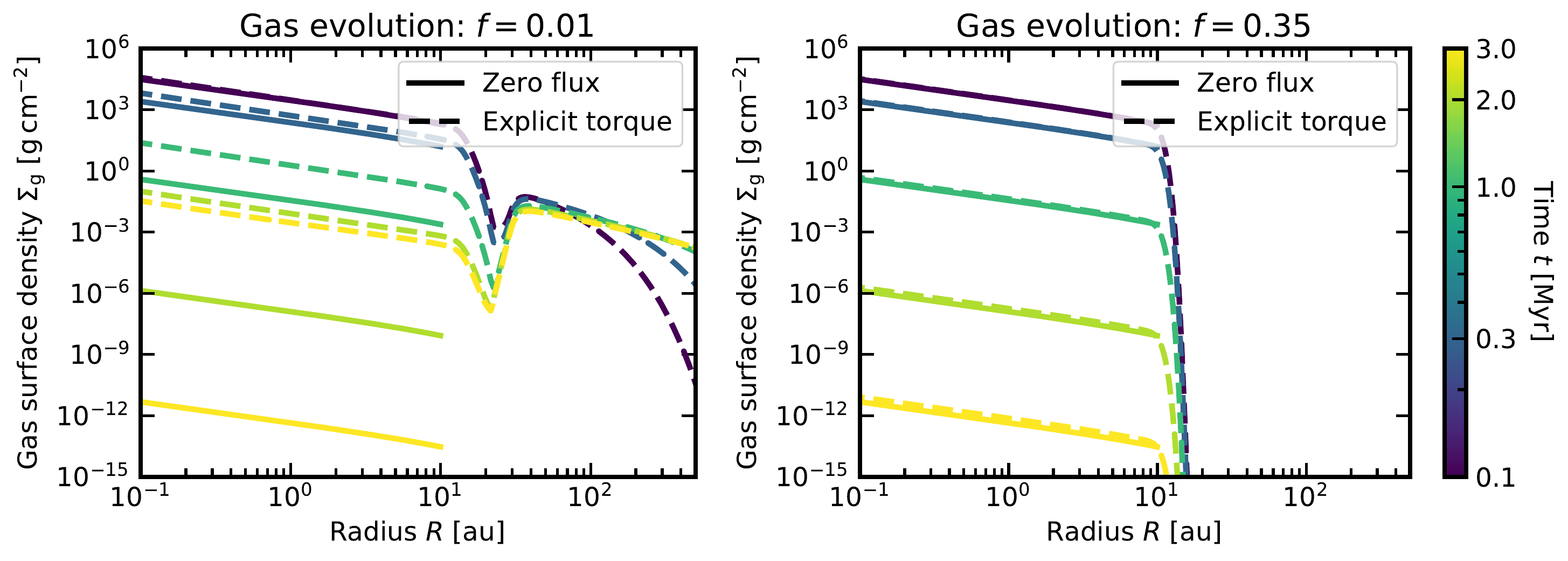}
    \caption{\textbf{Left panel:} Radial dependence of the disc gas surface density after $t=0.1,\,0.3,\,1,\,2\text{ and }3\text{ Myr}$. The solid lines identify the model enforcing the zero-flux boundary condition with $R_0=10\text{ au}$, $R_\text{trunc}=10\text{ au}$ and $\alpha=10^{-2}$. The explicit-torque solution with $R_0=10\text{ au}$, $q=0.11$, $R_\text{gap}=a/0.46=10\text{ au}$ and $\alpha=10^{-2}$ is plotted as a dashed line. \textbf{Right panel:} Same as in the upper panel with $f=0.35$.}
    \label{fig.A3}
\end{figure*}

Let us first focus on the $\alpha=10^{-3}$ case and discuss the differences in the gas and dust evolution between the two frameworks as a function of the expected truncation radius, $R_\text{trunc}$, and the binary mass ratio, $q$. In the widest binary model with $R_\text{trunc}=150\text{ au}$ both the gas and the dust surface density closely agree, regardless of the binary mass ratio. As the binary separation decrease, the explicit-torque and the zero-flux formulation show some discrepancies, which are mainly due to resolution effects. Once those are taken into account, also in the $R_\text{trunc}=10\text{ au}$ case both the gas and the dust surface density show no relevant differences for any value of $q$. 

As an example, in Fig.~\ref{fig.A2} the gas and the dust surface densities, $\Sigma_\text{g}$ and $\Sigma_\text{d}$, are plotted as a function of the disc radius after $t=0.1,\,0.3,\,1,\,2\text{ and }3\text{ Myr}$, in the left and right hand panels, respectively. The solid lines identify the zero-flux model snapshots with $R_\text{trunc}=10\text{ au}$, while the dashed ones refer to the explicit-torque model snapshots with $q=0.11$. The ratio between the gas masses in the explicit-torque and zero-flux formulation never exceeds a factor of 1.5. As for the dust masses, their ratio is always below 2.5.

Let us move on to the $\alpha=10^{-2}$ case. While no differences to the previous considerations apply in the case of wide binaries, for $R_\text{trunc}=10\text{ au}$ the evolution of the gas and the dust in the zero-flux and explicit-torque models differ substantially. This is clear from the left panel of Fig.~\ref{fig.A3} showing the gas surface density profiles as a function of the disc radius after $t=0.1,\,0.3,\,1,\,2\text{ and }3\text{ Myr}$. The solid lines identify the model enforcing the zero-flux boundary condition with $R_0=10\text{ au}$, $R_\text{trunc}=10\text{ au}$ and $\alpha=10^{-2}$. The explicit-torque solution with $R_0=10\text{ au}$, $q=0.11$, $R_\text{gap}=a/0.46=10\text{ au}$ and $\alpha=10^{-2}$ is plotted as a dashed line. 

Although this discrepancy can be reduced increasing the code resolution, it is generally true that the torque is not strong enough to efficiently open a deep gap and eventually truncate the disc. As a consequence, a relevant fraction of the gas expands over the gap, substantially increasing the disc lifetime. Moreover, the plot suggests that the gap deepens and widens on secular time scales, thus providing different disc dispersal velocities as time goes on. 

Carving a gap deep enough to prevent gas expansion is possible requiring that $f=0.35$. The right panel of Fig.~\ref{fig.A3} plots the radial dependence of the gas surface density after $t=0.1,\,0.3,\,1,\,2\text{ and }3\text{ Myr}$ using the same conventions as in the left panel, for a simulation enforcing $f=0.35$. Clearly now the explicit-torque and zero-flux solutions agree: the ratio between the gas masses in the explicit-torque and zero-flux formulation is always less than 1.5, as in the $\alpha=10^{-3}$ model with the same parameters.

We remark that in the $\alpha=10^{-2}$ and $R_\text{trunc}=10\text{ au}$ disc the dust surface density profiles in the explicit-torque and zero-flux models are substantially different whichever of the previous sets of parameters ($f$ and $q$) are enforced. In particular, more dust is retained in the explicit-torque scenario than in the zero-flux one. So, while in the gas it is possible to fine-tune the torque so as to avoid gas leakage beyond the gap and reconcile the predictions of the different torque implementations, more work is needed to address the differences arising in the dust.

To sum up, in wide binaries the explicit-torque and zero-flux formulations provide very close results. As the binary separation decreases and the disc viscosity increases, the agreement worsens. While in the gas it is always possible to modify the torque term to reconcile the two model predictions, the evolution of the dust needs to be analysed further.




\section{Is radial drift faster in binaries than in single-star discs?}\label{app:3}

\begin{figure*}
    \centering
	\includegraphics[width=1.5\columnwidth]{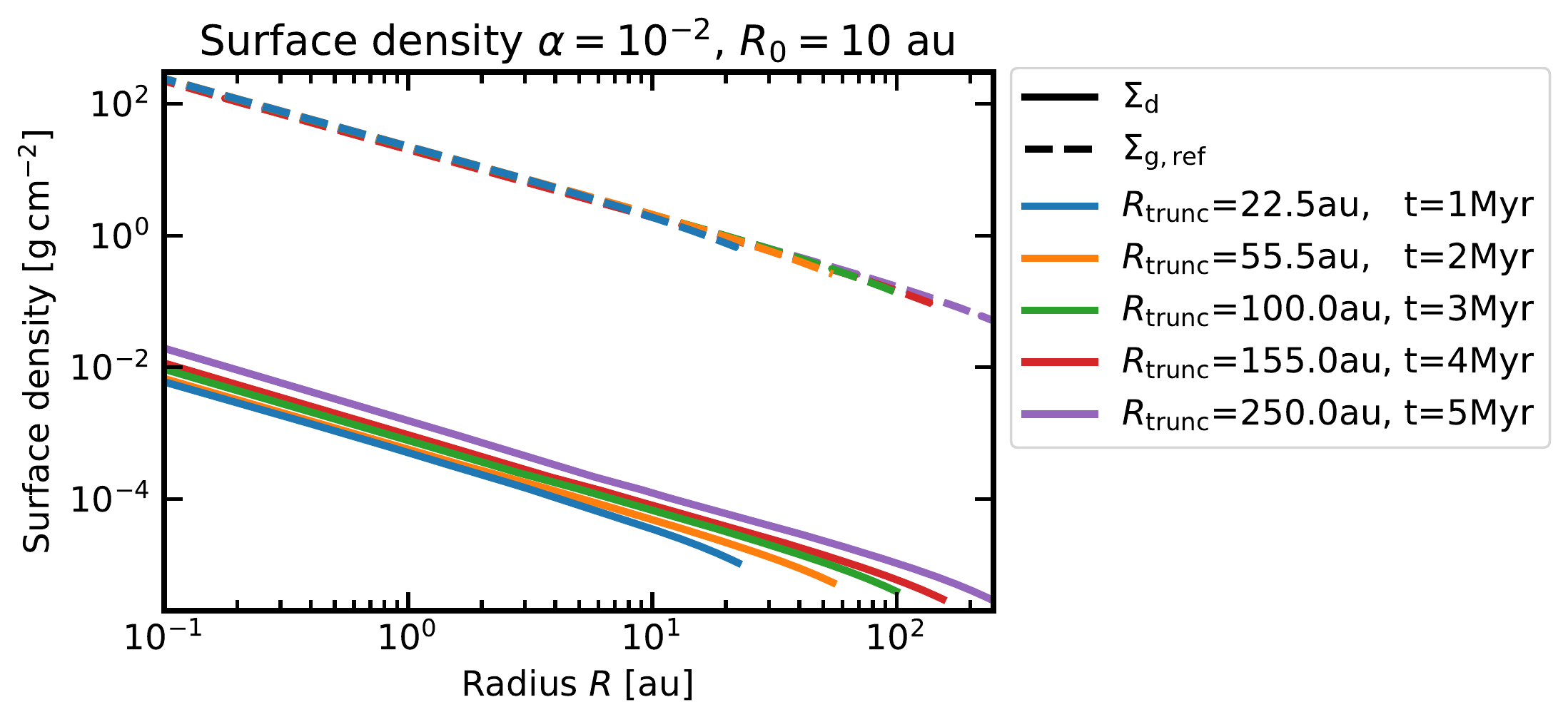}
    \caption{$\Sigma_\text{g,ref}$ as defined in the manuscript is plotted for $\alpha=10^{-2}$ and $R_0=10\text{ au}$. Binary discs with different $R_\text{trunc}$ are evolved until their final gas surface density - dashed lines - equals $\Sigma_\text{g,ref}$. The corresponding dust surface densities - solid lines - are also shown: different profiles of $\Sigma_\text{d}$ are associated with the same $\Sigma_\text{g,ref}$ as $R_\text{trunc}$ varies.}
    \label{fig.6.1}
\end{figure*}

In the body of this paper we proved that the presence of a stellar companion dramatically affects dust evolution in proto-planetary discs, enhancing the removal of dust grains due to radial drift. It would be interesting to assess if the faster dispersal of dust in binary discs is driven by the shorter time scale of the gas evolution or by an intrinsically faster removal of the dust grains, as a consequence of radial drift being more efficient than in single-star discs.

To answer this question, we study the evolution of the dust in binary discs with different truncation radii at the same stage of their  gas secular evolution: we do not compare different simulations \textit{at the same time}, but evolve them until they reach the \textit{same gas surface density}. Any differences in the integrated dust-to-gas ratio will be then due to the preferential removal of the dust with respect to the gas. For each set of initial parameters, $\alpha$ and $R_0$, we define the reference gas surface density, $\Sigma_\text{g,ref}$, as the gas surface density profile attained by the binary disc with $R_\text{trunc}=100\text{ au}$ after $t=3\text{ Myr}$. This choice is motivated by the evidence that almost all of our models generally prove to be drift-dominated when their gas surface density equals $\Sigma_\text{g,ref}$. We evolve a number of binary discs until their gas surface density equals $\Sigma_\text{g,ref}$, as shown in Fig.~\ref{fig.6.1} in the case of $\alpha=10^{-2}$ and $R_0=10\text{ au}$. Clearly, discs with smaller $R_\text{trunc}$ attain the reference gas surface density earlier.

\begin{figure*}
    \centering
	\includegraphics[width=\textwidth]{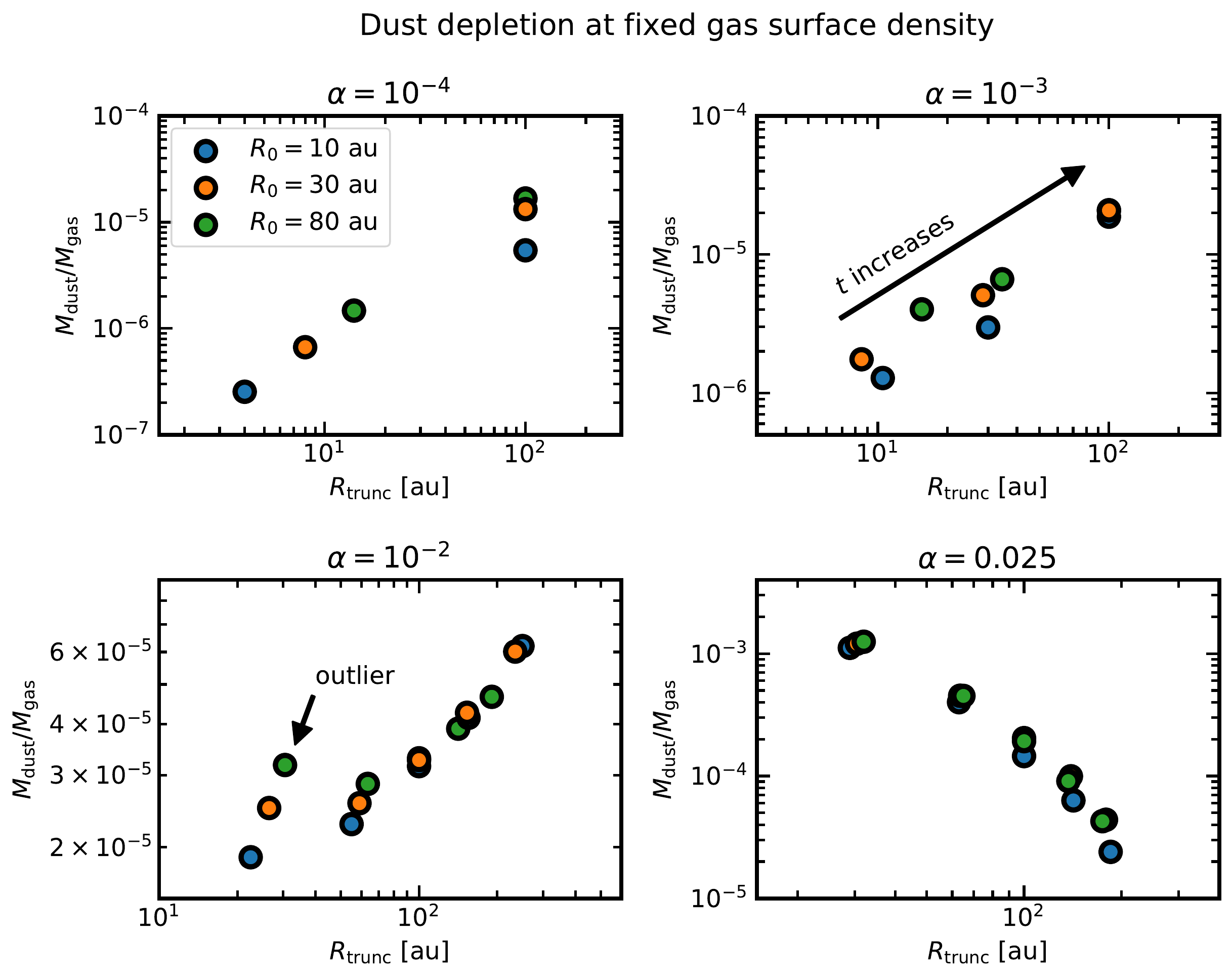}
    \caption{The integrated dust-to-gas ratio, $M_\text{dust}/M_\text{gas}$, is plotted as a function of the tidal truncation radius, $R_\text{trunc}$, for different values of $\alpha$ and the initial disc scale radius, $R_0$. All discs are evolved for different times, until they reach the same gas surface density.}
    \label{fig.6.2}
\end{figure*}

Fig.~\ref{fig.6.2} shows the dependence of the integrated dust-to-gas ratio on $R_\text{trunc}$, for different values of $\alpha$ and the initial disc scale radius, once $\Sigma_\text{g,ref}$ has been attained. Let us focus on the $\alpha=10^{-4}$ case. As it is clear from the figure, moving towards smaller truncation radii brings about a decrease in the integrated dust-to-gas ratio, \textit{despite smaller discs being significantly younger}. This proves that the faster dust removal in smaller discs can be attributed to an enhanced radial migration of the solids: because the time scale of radial drift grows with the distance from the star, the smaller $R_\text{trunc}$, the more effective the depletion. 
In the $\alpha=10^{-3}$ case, the behaviour is largely similar, with smaller discs losing dust faster.


Increasing the viscosity, larger regions of the model discs are in the fragmentation-dominated regime. Consequently, as dust drifts slower due to the small sizes imposed by fragmentation, the integrated dust-to-gas ratio is higher than for the lower viscosities. 
Although, the general evidence of smaller discs losing larger amounts of solids is still true, there is a notable exception to this trend: the disc with $R_0=80\text{ au}$ and $R_\text{trunc}=30.5\text{ au}$, which in Fig.~\ref{fig.6.2} is labelled as \textquotedblleft outlier\textquotedblright, retains more dust than its wider and older companions. Being this disc the shortest and youngest among those with the same initial scale radius, dust particles in the outer regions have grown the fastest, drifting earlier and accumulating in the inner fragmentation-dominated region. The pile-up of dust grains in the inner disc determines an enhancement of the integrated dust-to-gas ratio as not enough time has passed for those particles to be accreted with the gas flow. Interestingly, discs with comparably small tidal truncation radii but smaller scale radii retain less dust than their larger companion. Indeed, as the viscous time scale increases with $R_0$, more compact discs are more evolved: the dust had enough time to be accreted with the gas levelling off the accumulation of grains in the inner regions.

Finally, in the $\alpha=0.025$ case the behaviour of $M_\text{dust}/M_\text{gas}$ with $R_\text{trunc}$ has an opposite trend: the integrated dust-to-gas ratio decreases with $R_\text{trunc}$. At these viscosities discs are in the fragmentation-dominated regime throughout and dust removal is only due to gas accretion. The high value of the viscous parameter determines a fast depletion of gas, bringing about a substantial reduction of $a_\text{frag}$ with time. As the grain size cannot become smaller than the monomer grain size, $a_\text{min}=0.1\,\mu\text{m}$, as time goes on grain growth is inhibited out a characteristic radius. This is the location in the disc where the $a_\text{frag}=a_\text{min}$, which is determined only by the gas surface density at that time. By construction, as $\Sigma_\text{g,ref}$ is the same for every disc, also this characteristic radius is common to every set of models. Out of this radius grains drift efficiently. As a consequence, the larger $R_\text{trunc}$, the larger the drift-dominated region of the disc: dust grains in older discs, which are also those with larger $R_\text{trunc}$, experience considerable drift accumulating in the inner disc and increasing the local dust-to-gas ratio. As a result more dust can be accreted with the gas flow than in binary discs with a smaller tidal truncation radius. As a double-check, discs with lower $R_0$, whose gas accretion rate is higher, are poorer in dust than those with larger initial disc scale radius.

To sum up, we proved that in a large range of disc viscosities, namely $\alpha\lesssim10^{-2}$, the presence of a binary companion increases the efficiency of radial drift. In particular, the closer the binary the faster the grains drift. However, for larger viscosities the faster depletion of dust we witness in binary discs is mainly due to the shorter time scale of the gas evolution.


\bsp	

\label{lastpage}
\end{document}